\documentclass[a4paper,11pt]{article}
\pdfoutput=1 

\usepackage{jheppub} 
\usepackage[utf8]{inputenc}
\usepackage{slashed,cancel}
\usepackage{bm}
\usepackage{pifont}
\usepackage[english]{babel}
\usepackage{multirow}
\usepackage[normalem]{ulem} 
\usepackage{booktabs}
\usepackage{braket}
\usepackage{MnSymbol}
\usepackage{float}
\usepackage{graphicx}
\usepackage{subcaption}
\usepackage{multirow}
\usepackage[skins,theorems]{tcolorbox}
\usepackage[capitalise]{cleveref}
\usepackage{orcidlink} 

\numberwithin{equation}{section}

\preprint{IPPP/26/11}

\title{\boldmath Scalar Tsunamis from Black Hole Formation}

\author[a]{Arturo de Giorgi~\orcidlink{0000-0002-9260-5466}\,,}
\author[b]{Yeray Garcia del Castillo~\orcidlink{0009-0005-9159-205X}\,,}
 \author[c]{Joerg Jaeckel~\orcidlink{0000-0002-6038-4785}\,}

\affiliation[a]{Institute for Particle Physics Phenomenology, Department of Physics, Durham University, Durham DH1 3LE, U.K.}
\affiliation[b]{School of Physics, The University of New South Wales,
Sydney NSW 2052, Australia}
\affiliation[c]{Institut f\"ur Theoretische Physik, Universit\"at 
Heidelberg, Philosophenweg 16, 69120 Heidelberg, Germany}

\emailAdd{arturo.de-giorgi@durham.ac.uk}
\emailAdd{y.garcia\textunderscore del\textunderscore castillo@unsw.edu.au}
\emailAdd{jjaeckel@thphys.uni-heidelberg.de}

\abstract{Stars and other macroscopic objects may be surrounded by potentially large field configurations of very light scalars coupled to ordinary matter. If the star ends in a black hole, e.g. via a supernova or a neutron star merger, the source vanishes, and the field is released. In this paper, we improve on previous estimates~\cite{deGiorgi:2024pjb} for the field configurations arriving at large distances by including the effects of general relativity and an improved modelling of the initial field configurations. The total amount of energy released is typically of the same order of magnitude as suggested by simple flat space estimates. The spectrum receives noticeable corrections.}

\begin{document}
\maketitle
\flushbottom

%
%
%

\newpage

\section{Introduction}

Extreme astrophysical events such as the explosion of a star in a supernova or a neutron star merger provide a unique opportunity to search for new physics. 
In particular, in the search for new, feebly interacting particles, this has a long and important tradition. Notably, the supernova SN1987a was very quickly used to obtain stringent new constraints on axions, neutrino properties and more (cf., e.g.,~\cite{Raffelt:1987yt,Turner:1987by,Burrows:1988ah,Grifols:1996id,Raffelt:1996wa,Longo:1987ub,Stodolsky:1987vd,Grifols:1988ff}) and continues to be employed for this purpose  (see~\cite{Brockway:1996yr,Giannotti:2010ty,Payez:2014xsa,Jaeckel:2017tud,Fiorillo:2022cdq,Diamond:2023scc,Manzari:2023gkt,Fiorillo:2025sln,Manzari:2025jbc,Ferreira:2025qui} for just a few examples). Most of these works utilize the high temperatures and densities that are present for short amounts of time for copious particle production. This then leads to either directly visible production or an excessive energy loss that would leave traces in other observables.

In this note, we want to further develop a somewhat different route, outlined in~\cite{deGiorgi:2024pjb}. There, we considered a situation where a sizeable field configuration of a new, nearly massless field, surrounding the initial object, is present before the event.\footnote{See e.g. also~\cite{Yoshino:2012kn,Eby:2016cnq,Levkov:2016rkk,Baumann:2018vus} for the generation of transient signals generated in cosmic events with initial field configurations and~\cite{Fiorillo:2025gnd,Lecce:2025dbz,Lecce:2025vjc,Candon:2025ypl} for some recent examples of transient signals via particle production.}  The event can be, e.g., a supernova or a binary neutron star merger that leads to black hole~(BH) formation. The ordinary matter (mostly) ends up in the black hole. As the field that was originally sourced by the coupling to matter, this coupling/sourcing vanishes, the field is released, and a field ``Tsunami'' propagates outwards from the source. Travelling to Earth it then can cause transient signals in experiments \cite{PhysRevD.99.082001,GNOME:2023rpz,DMRadio:2022pkf,Shaw:2021gnp,DMRadio:2023igr,Sun:2024qis}. 

In Ref.~\cite{deGiorgi:2024pjb}, we estimated the field evolution and resulting signals in terrestrial experiments by simply taking a free field evolution in flat space, neglecting the BH background. This is clearly a very simplistic assumption. In the present work, we now want to take the next step and include the effects of the black hole in the evolution of the scalar field configuration. To do this, we start with the original field configuration and evolve it in the black hole background metric to determine the signal as arriving at large distances. We study both the total amount of energy released as well as the spectrum, comparing them to those obtained from the simple flat space estimate.
While the energy released is at a similar level, we find notable changes in the spectrum.

The process of BH formation may also change the initial field configuration, in turn affecting the signal.
Therefore, we also consider additional field configurations that, while still being relatively simple, reflect aspects of a pre-black hole formation collapse of ordinary matter. Again, we find moderate quantitative changes.

The remainder of the paper contains the following steps. In Sec.~\ref{sec:initialfields} we introduce the setup and, in particular, the initial field configurations we will use. This is followed in Sec.~\ref{sec:grevolution} by a discussion of how we evolve the field in the gravitational field of the black hole and how we calculate the relevant energies, both in principle and numerically. 
Results for the various initial conditions are presented in Sec.~\ref{sec:results} with further discussion and some conclusions being provided in Sec.~\ref{sec:conclusions}. We briefly remark on some aspects of how the scalar field evolution in our approximation relates to the black hole no-hair theorem~\cite{Israel:1967wq,Israel:1967za,Carter:1971zc,Ruffini:1971bza,Carter:2009nex,Robinson:1975bv,Carter:1979wef,Mazur:2000pn} in Appendix~\ref{app:nohair}.

\section{Model and initial field configurations}
\label{sec:initialfields}

\subsection{Model}
Our model is as simple as it gets.
It is a real, essentially massless scalar field $\phi$.
Indeed, as discussed in~\cite{deGiorgi:2024pjb} a mass 
\begin{equation}
    m_{\phi}\lesssim 1/r\lesssim 10^{-25}\,{\rm eV},
\end{equation}
is required to avoid signal dispersion over typical distances from astrophysical objects to Earth, which would destroy transient signatures.

Accordingly, in practice, we consider a massless field, minimally coupled to gravity via the metric $g_{\mu\nu}$
\begin{equation}
    \mathcal{L}=g^{\mu\nu}\frac{1}{2}\partial_\mu \phi\partial_\nu\phi\,.
\end{equation}
Interactions with the source as well as self-interactions relevant before BH formation are taken to be modelled by the initial conditions described in the following Sec.~\ref{sec:initial-conditions}.
We also neglect further couplings in the evolution of the field shortly after the BH formation. However, they might be relevant for a faithful description of the propagation over large distances. Indeed, small interactions of the scalar field wave with local matter density or magnetic fields (depending on the couplings) can have sizeable impact on the wave's propagation once integrated over large distances~\cite{Arakawa:2025hcn}.

\subsection{Static initial conditions}
\label{sec:initial-conditions}
\begin{figure}
    \centering
    \begin{subfigure}[]{0.48\textwidth}
        \includegraphics[width=\linewidth]{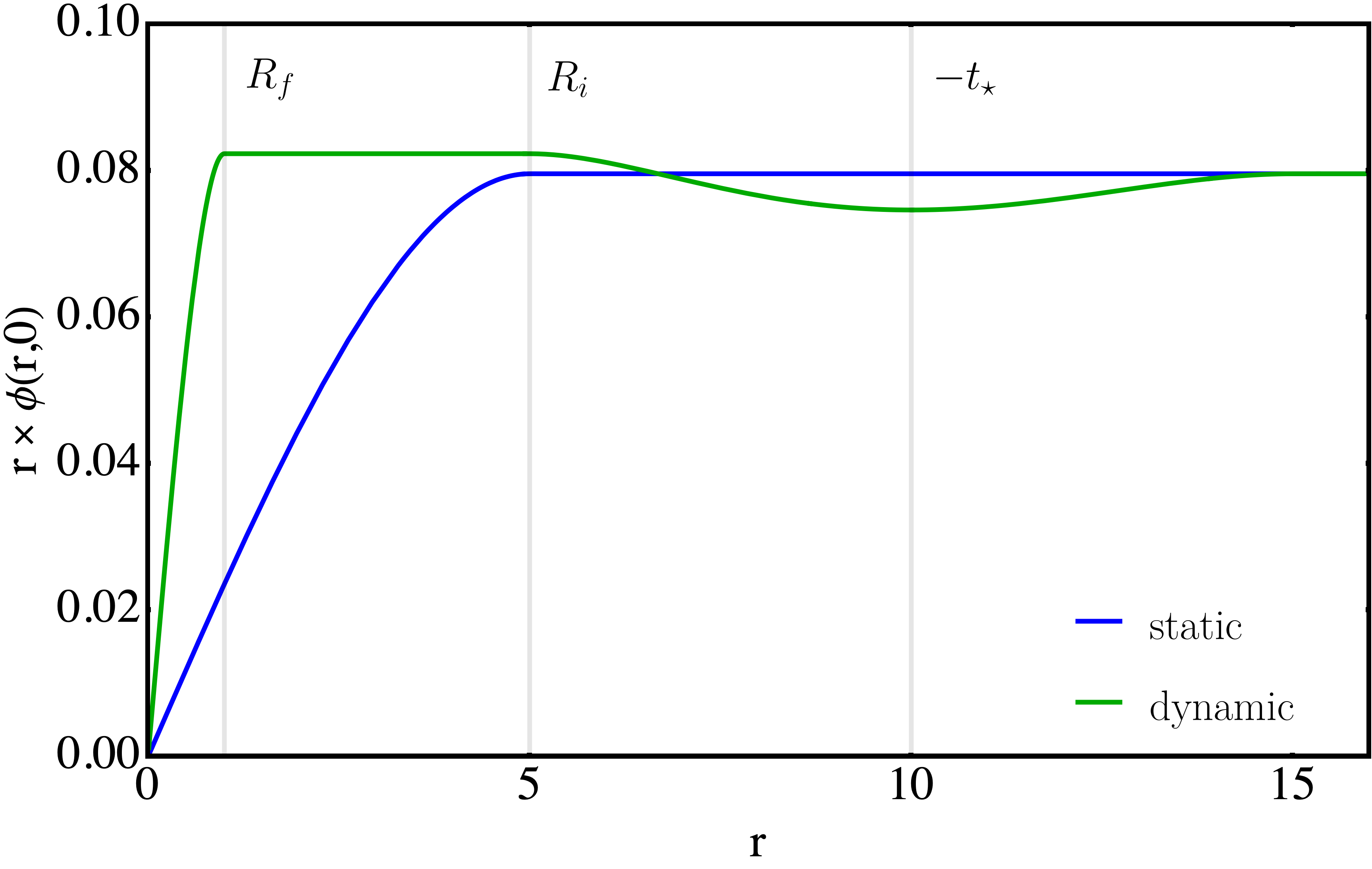}
        \caption{Field profile at $t=0$.}
    \end{subfigure}
    \begin{subfigure}[]{0.5\textwidth}
        \includegraphics[width=\linewidth]{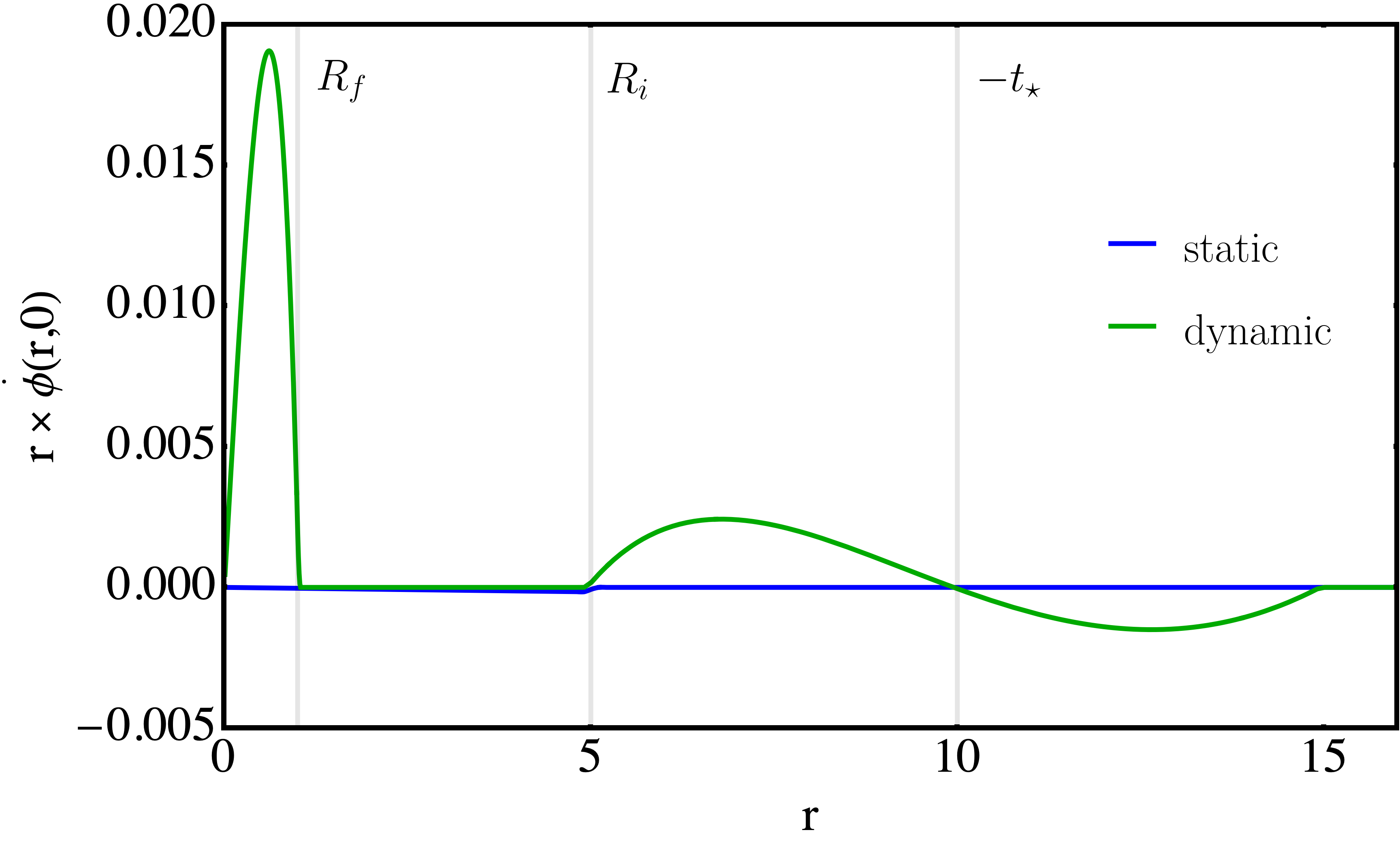}
        \caption{Time derivative at $t=0$.\label{fig:initial2}}
    \end{subfigure}
    \caption{Example of static (blue) and dynamical (green) initial conditions for the Yukawa-like case in arbitrary units. The left (right) plot shows the field value (time derivative) at $t=0$, when the BH forms. The collapse velocity was taken to be $v=0.4$, while the initial and final radii are taken to be $R_i=5$, $R_f=1$; the collapse then begins at $t_\star=-10$.}
    \label{fig:initial}
\end{figure}

As simple examples, we examine the following spherically symmetric initial field configurations, some of which were also used in~\cite{deGiorgi:2024pjb}, 
\begin{enumerate}
    \item A Yukawa-like profile:
    \begin{equation}
        \label{eq:initial-Y}
        \phi_{Y,0}(r)=\frac{g_Y}{4\pi r}\,.
    \end{equation}
    Such a configuration can be generated simply by a Yukawa coupling of the field to the matter density of the BH progenitor.
    In a slightly more sophisticated version, we can use a homogeneous charge distribution in the progenitor of radius $R$, such that
    \begin{align}
     &     \phi_{Y,0}(r) =\dfrac{g_\text{Y}}{4\pi}\times\begin{cases}
     \dfrac{1}{2 R}\left(3-\dfrac{r^2}{R^2}\right)\,, & r\leq R\,,\\
         \dfrac{1}{r}\,, & r> R\,.
     \end{cases}
    \end{align}
    The initial field profile is depicted in Fig.~\ref{fig:initial}.
    \item A ``compact" profile:
    \begin{equation}
        \phi_{C,0}(r)=\begin{cases}
            \label{eq:initial-C} g_C \frac{(R-r)^3}{R^4}\,, &r\leq R\,,\\
            0\,, & r>R\,.
        \end{cases}
    \end{equation}

    This configuration models a situation where non-linear interactions (e.g. some form of screening mechanism~\cite{Khoury:2003rn,Hook:2017psm,Balkin:2021zfd,Balkin:2023xtr}) cluster the field or at least a sizeable part of it more densely in the initial object.
    \item A spherical shell:
    \begin{align}
    \phi_{\text{S},0}(r) =
    \begin{cases}
    0\,, & r \le R_1, \\
    \frac{g_{S}}{r}\frac{ (r - R_1)^3 (R_2 - r)^3}{(R_2-R_1)^6}\,, & R_1 < r < R_2, \\
    0\,, & r \ge R_2\,.
    \end{cases}
    \end{align}
    This case is perhaps the least realistic to occur in Nature. However, due to the possibility of precisely localising the field energy content with respect to the BH's horizon, it provides a pedagogical case to understand how energy gets reflected or absorbed by the BH.
\end{enumerate}

We will use configurations 1 and 2 for a direct comparison with the flat space dynamics obtained in Ref.~\cite{deGiorgi:2024pjb}.

\subsection{Dynamical initial conditions}
In supernovae, the situation is quite dynamic already before black hole formation. One can therefore wonder what happens to the scalar field configuration during this time. This may then set a different initial condition for the later evolution in a black hole background.
To get an idea of the size of these effects, we consider the following modifications of the static initial conditions discussed above.
\begin{enumerate}
\item{} As an extreme case, we simply take the static initial field configuration, but for the evolution, we take only the inward-moving part of the spectrum, i.e. those that move towards the black hole at the speed of light. Let us parametrise $\phi(r)=\psi(r)/r$; then, the in-going wave condition sets $\psi(r,t)\sim \exp[i(r+t)]$, and thus
\begin{align} \nonumber
 \dot\psi_\text{IC}(r) &=\psi'_\text{IC}(r)~.
\end{align}
\item{} We calculate the field configuration (in flat space) of a homogeneously charged sphere that contracts with a velocity $v$.
This is motivated by a simple picture of the supernova collapse. Notably, simulations of supernovae find a typical contraction velocity in the semi-non-relativistic regime $v\lesssim 0.5$~\cite{Kuroda:2023mzi}.

For a Yukawa-like case, we model the charge density sourcing the scalar field as
\begin{equation}
\label{eq:collapse1}
    J(r,t)=\dfrac{3g_Y}{4\pi R(t)^3}\Theta(-t)\Theta(R(t)-r)\,,
\end{equation}
with
\begin{equation}
\label{eq:collapse2}
    R(t)=R_i+\Theta(t-t_\star)\left(\frac{R_f-R_i}{-t_\star}\right)\left(t-t_\star\right)\equiv R_i-\Theta(t-t_\star)v\left(t-t_\star\right)\,,
\end{equation}
such that it begins contracting at some $t_\star <0$ and collapses into a BH at $t=0$. More details about the field evolution with such initial conditions can be found in App.~\ref{sec:contratingsource}.
\end{enumerate}

A comparison between the static and dynamical initial conditions for the Yukawa-like case can be seen in Fig.~\ref{fig:initial}. The parameters of the dynamical initial condition were chosen to fix the collapse velocity to $v=0.4$: the initial and final radii are taken to be $R_i=5$, $R_f=1$ and the collapse begins at $t_\star=-10$ in arbitrary units.
The field wave centred at $r=-t_\star$ corresponds to the emission when the source begins its contraction. Due to causality, it extends to $r=-t_\star\pm R_i$. After the initial kick that starts the contraction, the collapse is linear in time, with null acceleration, and thus no energy is emitted. Finally, at $t=0$ the source vanishes, and the field waves get emitted; again, due to causality, the field wave extends to $r=R_f$. 

We note that this is, of course, still a very simple description of the contraction during a supernova. In particular, in a dynamical model that also includes an acceleration of the source distribution, $\ddot{R}(t)$, there is a part of the field emission that arises from this acceleration and not from the disappearance of the source at $t=0$. In our chosen model and parameters, this corresponds to the non-vanishing field values at $r\geq 5$ in Fig.~\ref{fig:initial2}. We think that this part is likely more model-dependent.


\section{Scalar field evolution in the black hole background}
\label{sec:grevolution}

\subsection{Equations of motion and the black hole potential}
We work in a static black hole~(BH) background described by the Schwarzschild metric, c.f., e.g.~\cite{Schwarzschild:1916uq,Misner:1973prb},
\begin{align}
\label{eq:metric}
    &ds^2=f(r)dt^2-f(r)^{-1}dr^2-r^2 d\Omega^2\,, &f(r)=1-\frac{r_H}{r}\,,
\end{align}
where $r_H=2G_N M$ is the BH's horizon radius, $M$ is the BH mass, and $G_N$ is Newton's constant.
The equation of motion~(EOM) of a free massless scalar field in a generic background is
\begin{equation}
    0=\sqrt{-g}\,\phi_{;\mu}^{;\mu}=\partial_\mu(\sqrt{-g}g^{\mu\nu}\phi_{,\nu})\,.
\end{equation}
If the field is spherically symmetric, $\phi(r,t)$, the EOM simplifies to
\begin{equation}
    \label{eq:original-EOM}\Ddot{\phi}-\frac{f(r)}{r^2}\frac{d}{dr}\left[r^2 f(r) \frac{d\phi}{dr}\right]=0\,,
\end{equation}
where $\dot{\phi}$ and $\phi'$ are time and spatial derivatives, respectively.

As often in GR, the main challenge is to choose suitable coordinates for the studied problem. The coordinates defined above, while having an intuitive interpretation, are not the optimal choice for our purpose. The numerical simulations quickly meet the apparent horizon's singularity, making the simulations unstable. Therefore, we choose \textit{tortoise} coordinates defined by
\begin{align}
    \label{eq:tortoise}&\frac{dx}{dr}\equiv\frac{1}{f(r)}\,, &x=r+r_H\log\left|\frac{r}{r_H}-1\right|\,,
\end{align}
The above coordinates have two advantages. First, the effect of the curvature is distributed equally to the time and the radial part of the metric, thus making some of the expressions simpler,
\begin{equation}
    ds^2=f(x)(dt^2-dx^2)-r(x)^2d\Omega^2\,.
\end{equation}
Secondly, they move the horizon to $x\to -\infty$. This greatly simplifies all numerical calculations, as divergences associated with the horizon are always removed due to the finite numerical domain of the simulation.

\begin{figure}
    \centering
    \includegraphics[width=0.6\linewidth]{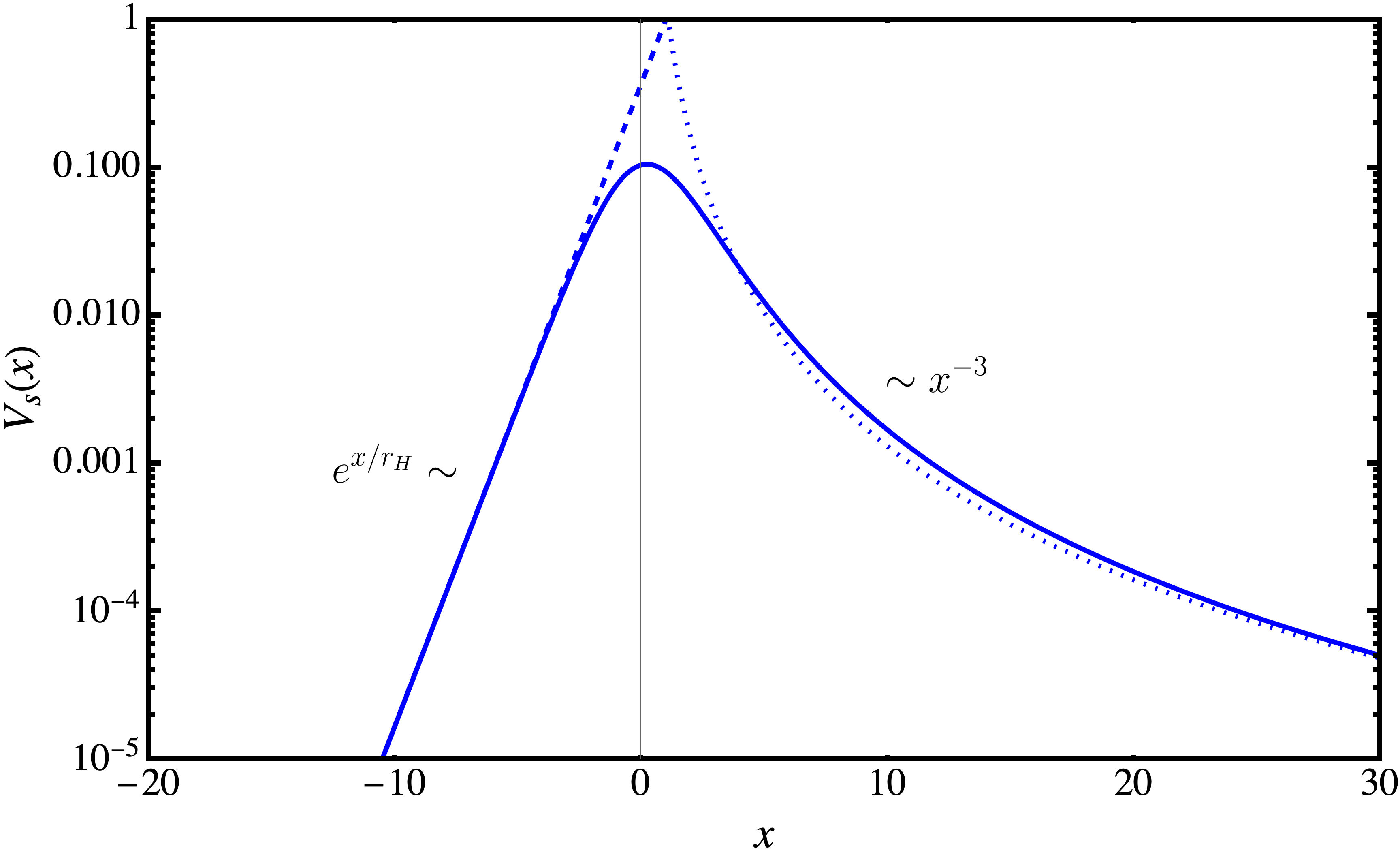}
    \caption{BH potential in Tortoise coordinates for $r_H=1$ in arbitrary units.}
    \label{fig:potential}
\end{figure}

Let us now consider the EOM~\eqref{eq:original-EOM} in the new coordinates. By employing the ansatz
\begin{equation}
    \label{eq:ansatz}\phi(r,t)=\frac{\psi(r)}{r}\,,
\end{equation}
we obtain the wave-like equation of motion
\begin{align}
   &\Ddot{\psi} = \frac{d^2\psi(x)}{dx^2}-V_s(x)\psi(x)\,, &V_s\equiv \frac{f(r)f'(r)}{r}=\left(1-\frac{r_H}{r(x)}\right)\frac{r_H}{r(x)^3}\,.
   \label{eq:eomtortoisse}
\end{align}
The above equation describes a wave propagating at the speed of light subject to an external potential $V_s(x)$. 
The effect of the curved background is entirely captured by the potential $V_s$, whose profile is plotted in Fig.~\ref{fig:potential}. 

Let us briefly examine some of its properties.
The potential is peaked around $x=0$,
\begin{align}
    &r_\text{peak}= \frac{4}{3}r_H \,, && x_\text{peak}=\left(\frac{4}{3}-\log(3)\right)\times r_H\approx 0.23~r_H\,, &&V_{s,\text{peak}}=\frac{27}{256}\left(\frac{1}{r_H^2}\right)\approx \frac{0.105}{r_H^2}\,.
\end{align}
For larger Schwarzschild radii, the potential barrier gets wider, but its peak value gets smaller.
In tortoise coordinates, the horizon is at $x=-\infty$, while for $r\gg r_H$, the two sets of coordinates match
\begin{align}
    &r\to r_H \, \Rightarrow x\to- \infty\,,&r\to +\infty \, \Rightarrow x\to+ \infty\,.
\end{align}
The asymptotic behaviour can be extracted straightforwardly. Far from the BH, $r\approx x$, and thus
\begin{equation}
    V_s(r\gg r_H)\sim \frac{1}{x^3}\,.
\end{equation}
Near the BH, one has
\begin{align}
    &x(r\sim r_H)\approx r_H\log\left(\frac{r}{r_H}-1\right)\,, &r\approx r_H\left(1+e^{(x-r_H)/r_H}\right)\,,
\end{align}
and consequently
\begin{align}
    &f(x\to-\infty)\approx e^{x/r_H}\,, &V_s(r\sim r_H)\sim e^{x/r_H}\,.
\end{align}

An exact closed solution to Eq.~\eqref{eq:eomtortoisse} does not exist, and much work has been done in the literature to extract salient features analytically. In the following, we will adopt a numerical approach.

\bigskip

That said, before proceeding with the description of the numerical method employed in this work, it is worth describing some of the salient, known~\cite{Ferrari:1981dh,Blome:1981azp,Ferrari:1984ozr,Schutz:1985km,Iyer:1986np,Konoplya:2004ip,Price:1971fb,Ching:1995tj,Donninger:2009rc,Donninger:2009zf} physics features of a scalar field propagating in a BH background. 
After the BH formation, the scalar field becomes dynamical and starts propagating both inwards and outwards. The potential barrier of Eq.~\eqref{eq:eomtortoisse} and sketched in Fig.~\ref{fig:potential} has multiple non-trivial effects on the field's evolution. Schematically, one can qualitatively distinguish three regimes:
\begin{enumerate}
    \item Initial 
    transmission/reflection of the field through/at the potential barrier. The net effect strongly depends on the initial field configuration. In the most extreme cases, where the initial field configuration is mainly localised far away from the horizon, the BH can reflect almost the entirety of the incoming wave. This is because the low-frequency modes have a more difficult time penetrating the barrier.
    \item A trapping regime caused by the repeated backscattering of the wave off the effective potential. In time, energy continuously leaks both into the horizon and out to infinity. The leakage turns the field evolution near the horizon into a damped “ringdown”, whose time dependence for each mode is well approximated by a complex frequency $\omega =\omega_R-i\omega_I$ such that
    \begin{equation}
        \phi(t)\propto e^{-i\omega t}=e^{-\omega_I t}\times e^{-i\omega_R t}\,.
    \end{equation}
    Because the field must satisfy purely ingoing boundary conditions at the event horizon and purely outgoing boundary conditions at infinity, only a discrete set of complex frequencies is allowed. These discrete damped resonances are usually referred to as the BH quasi‑normal modes~(QNMs) and have been intensively studied in the literature~\cite{Ferrari:1981dh,Blome:1981azp,Ferrari:1984ozr,Schutz:1985km,Iyer:1986np,Konoplya:2004ip}. They depend only on the BH geometry and on the spin and angular momentum of the perturbation.
    For example, for a scalar field, the lowest s-wave QNM angular frequency is given by~\cite{Konoplya:2004ip}
    \begin{equation}
        \omega \approx \left(\frac{2}{ r_H}\right)\left(0.1105 - 0.1049i\right)\,.
    \end{equation}
    \item A late-time power-law field tail produced by backscattering off the long-range part of the curvature potential. This characteristic behaviour goes in the literature under the name of ``Price’s law"~\cite{Price:1971fb,Ching:1995tj,Donninger:2009rc,Donninger:2009zf}. The exponent of the power-law depends on the initial configuration of the perturbation. For an initially static scalar Yukawa-like and compact distribution, the field decays asymptotically as $t^{-3}$ and $t^{-4}$, respectively.
\end{enumerate}
We checked the compatibility between the scaling of the numerical results and the expectations of Price's law to validate the faithfulness of our code against independent, known results~\cite{Price:1971fb,Ching:1995tj,Donninger:2009rc,Donninger:2009zf}. Some examples can be found in Fig.~\ref{fig:plotasympcomb} of Appendix~\ref{app:extrafig}. As we will see in our discussion below, we also find the expected~\cite{Berti:2009kk,Kokkotas:1999bd} first quasi-normal mode providing another check of our calculation.

\subsection{Numerical solution in tortoise coordinates}
In order to solve Eq.~\eqref{eq:eomtortoisse} numerically, it is useful to define $\hat{r}=r/r_H$ and $\hat{t}=t/r_H$ which yields the dimensionless equation
\begin{equation}
   \frac{d^2\psi(\hat{x})}{d\hat{t}^2} = \frac{d^2\psi(\hat{x})}{d\hat{x}^2}-V_s(\hat{x})\psi(\hat{x})\,,
   \label{eq:motion}
\end{equation}
with $\hat{x}=\hat{r}+\log\left|\hat{r}-1\right|$. This equation can be solved numerically for a finite interval $\hat{r}\in[1+\epsilon,\hat{r}_{\text{max}}]$ which corresponds to $\hat{x}\in[\hat{x}_{\text{min}},\hat{x}_{\text{max}}]$. Since $\hat{r}(\hat{x})$ cannot be found analytically, we invert Eq.~\eqref{eq:tortoise} numerically in order to find this interval and the potential $V_s(\hat{x})$ explicitly. 

We solve the PDE on a finite interval, $\hat{x}_{\text{min}}< \hat{x} < \hat{x}_{\text{max}} $, with the initial state specified as  
\begin{equation}
   \psi( \hat{t}=0, \hat{x}) =  \psi_{\mathrm{IC}}(\hat{x}), 
\qquad 
\dot\psi( \hat{t}=0, \hat{x}) =  \dot\psi_{\mathrm{IC}}(\hat{x}), 
\end{equation}
and subject to the boundary conditions
\begin{align} 
 &\psi( \hat{t}, \hat{x}_\text{min}) = \psi_{\mathrm{IC}}(\hat{x}_\text{min})~,
&\psi( \hat{t}, \hat{x}_\text{max}) = \psi_{\mathrm{IC}}(\hat{x}_\text{max})~.
\label{boundary-conditions-numerical}
\end{align}
As we discuss below, we choose our simulation time such that there are no effects of these boundary conditions within the range of  interest in $\hat{x}$.

The initial profiles $\psi_{\mathrm{IC}}, \dot\psi_{\mathrm{IC}}$ are obtained from the initial field configurations discussed in Sec.~\ref{sec:initialfields}.

To carry out the numerical solution, we use the standard \emph{method of lines}, employing the implementation already used in~ \cite{Burrage:2024mxn}. The spatial domain is divided into $N$ intervals of size ${d\hat{x}} = (\hat{x}_\text{max}-\hat{x}_\text{min})/N$. Spatial derivatives are replaced by second-order finite-difference operators, which convert the PDE into a coupled system of $(N-2)$ second-order ODEs. The two missing equations come directly from the boundary conditions. The resulting ODE system is then integrated using a standard time-evolution solver.  

The boundary conditions in Eq.~\eqref{boundary-conditions-numerical} have been chosen in order to be consistent with the initial field configuration. Note, however, that, in principle, one would expect the field to evolve to different values at these points and, imposing these boundary conditions, may lead to an unphysical behaviour. Yet, our focus lies primarily on the transient dynamics of the system, meaning we do not need to compute the field evolution indefinitely. We note that, Eq.~\eqref{eq:motion} corresponds to a wave equation that propagates at the speed of light and the potential is negligible at the boundaries. Hence, if we choose $\hat{x}_\text{min}\ll0$ and $\hat{x}_\text{max}\gg0$, we do not expect any noticeable boundary effects as long as the running time fulfils $\hat{t}\lesssim\text{min}(|\hat{x}_\text{min}|,|\hat{x}_\text{max}|)$. Additionally, one needs to choose a region large enough to include the features of the initial configuration within the simulation.
One also has to be careful in choosing $d\hat{x}$ appropriately. The characteristic length scale of the potential is given by $r_H$, so we should use a $d\hat{x}$ which can resolve this, which in dimensionless variables corresponds to ($d\hat{x}<1)$. However, the initial field configuration may introduce an additional characteristic length scale $\hat{\lambda}$, which also needs to be resolved. Therefore, $d\hat{x}$ should be chosen to satisfy $d\hat{x} \lesssim \min \bigl(1, \hat{\lambda}\bigr)$.

\subsection{Field energy in a black hole background}
A simple determinant for the size of the observable signal is the total energy that is released. We therefore want to compare this quantity in the simplistic flat space approximation to the calculation including the BH background. 
To do this, it is, however, important to fairly compare the released energy to the initial energy stored in the field configuration, including the effect of the non-trivial metric.
Therefore, let us recall a suitable energy measure.
As a small tangent, we comment on the energy flow of the field in the presence of a BH background and the behaviour of the horizon in Appendix~\ref{app:nohair}.

The Hamiltonian density of the field in spherical coordinates is given by
\begin{equation}
    \mathcal{H}=\sqrt{-g}\left[\, \frac{g^{00}}{2}\dot{\phi}^2-\frac{g^{11}}{2}(\phi')^2\right]\,.
\end{equation}
Here, the spatial derivative is taken with respect to $r$ or $x$, depending on the choice of coordinates. Specialising to the tortoise case, the Hamiltonian density reads
\begin{equation}
    H=\int_V  \mathcal{H}=(2\pi)\int\limits_{-\infty}^\infty dx\, \,\left(\dot{\psi}^2+\left(\psi'-\psi \frac{f(x)}{r(x)}\right)^2\right)\,.
    \label{eq:hamiltonian}
\end{equation}
Additional details on the derivation can be found in App.~\ref{app:energy}.

Turning to the energy stored in the field. In Ref.~\cite{deGiorgi:2024pjb}, the energy of the field was calculated under the assumption of a flat metric. Here, we go beyond this simplistic approximation. 

Let us consider an object that is approximately a BH but not quite yet. We will denote by $E_f$ and $E_c$ the energies in a flat and curved background, respectively. Since the energy within the will-be horizon is lost, we include only the energy stored from $r\in(r_H,\infty)$, 
\begin{align}
    &E_f=(2\pi)\int\limits_{r_H}^\infty dr\, r^2\,\left(\dot{\phi}^2+(\partial_r\phi)^2\right)=(2\pi)\int\limits_{-\infty}^\infty dx\, r(x)^2 f(x)\,\left(\dot{\phi}^2+\frac{1}{f(x)^2}(\partial_x\phi)^2\right)\,,\\ &E_c=(2\pi)\int\limits_{r_H}^\infty dr\, r^2 f(r)\,\left(\frac{\dot{\phi}^2}{f(r)^2}+(\partial_r\phi)^2\right)=(2\pi)\int\limits_{-\infty}^\infty dx\, r(x)^2\,\left(\dot{\phi}^2+(\partial_x\phi)^2\right)\,. 
\end{align}

With this, we can compute the differences between the flat (f) and curved (c) cases. Employing the profiles defined in Eqs.~\eqref{eq:initial-Y}-\eqref{eq:initial-C}, the energies for an initially static configuration, $\dot{\phi}=0$, read
\begin{align}
    &E_{Y,f}=\frac{g_Y^2}{8\pi r_H}\,, &&E_{C,f}=g_C^2~\frac{6 \pi  }{35 r_H z^8}(z-1)^5 \left[z (z+5)+15\right]\,,\\
    &E_{Y,c}=\frac{g_Y^2}{16\pi r_H}\,,  &&E_{C,c}=g_C^2~\frac{3 \pi  }{35 r_H z^8}(z-1)^6 (2 z+5)
\end{align}
where we assumed $R> r_H$ and $R\equiv r_H z$\,.

The energies between the flat and curved cases differ by a factor of $2$ for the Yukawa initial profile and for the Compact case by
\begin{equation}
    \frac{E_{C,c}}{E_{C,f}}\approx \begin{cases}
        1-\frac{7}{2z}\,, & R\gg r_H~(z\gg 1)\,,\\
        \frac{z-1}{6}\,, & R\approx r_H~(z\approx 1)\,.
    \end{cases}
\end{equation}
Due to the attractive nature of gravity the energy in curved spacetime is always smaller and approaches its flat value only when the radius of the initial source is much larger than the BH's horizon. If the radius is comparable to the horizon size, the energy gap can be sizeable, e.g. for $z=2$ one finds $E_{C,c}/E_{Y,c}\approx 1/6$.

\begin{figure}
    \centering
    \begin{subfigure}[]{0.495\textwidth}
        \includegraphics[width=\linewidth]{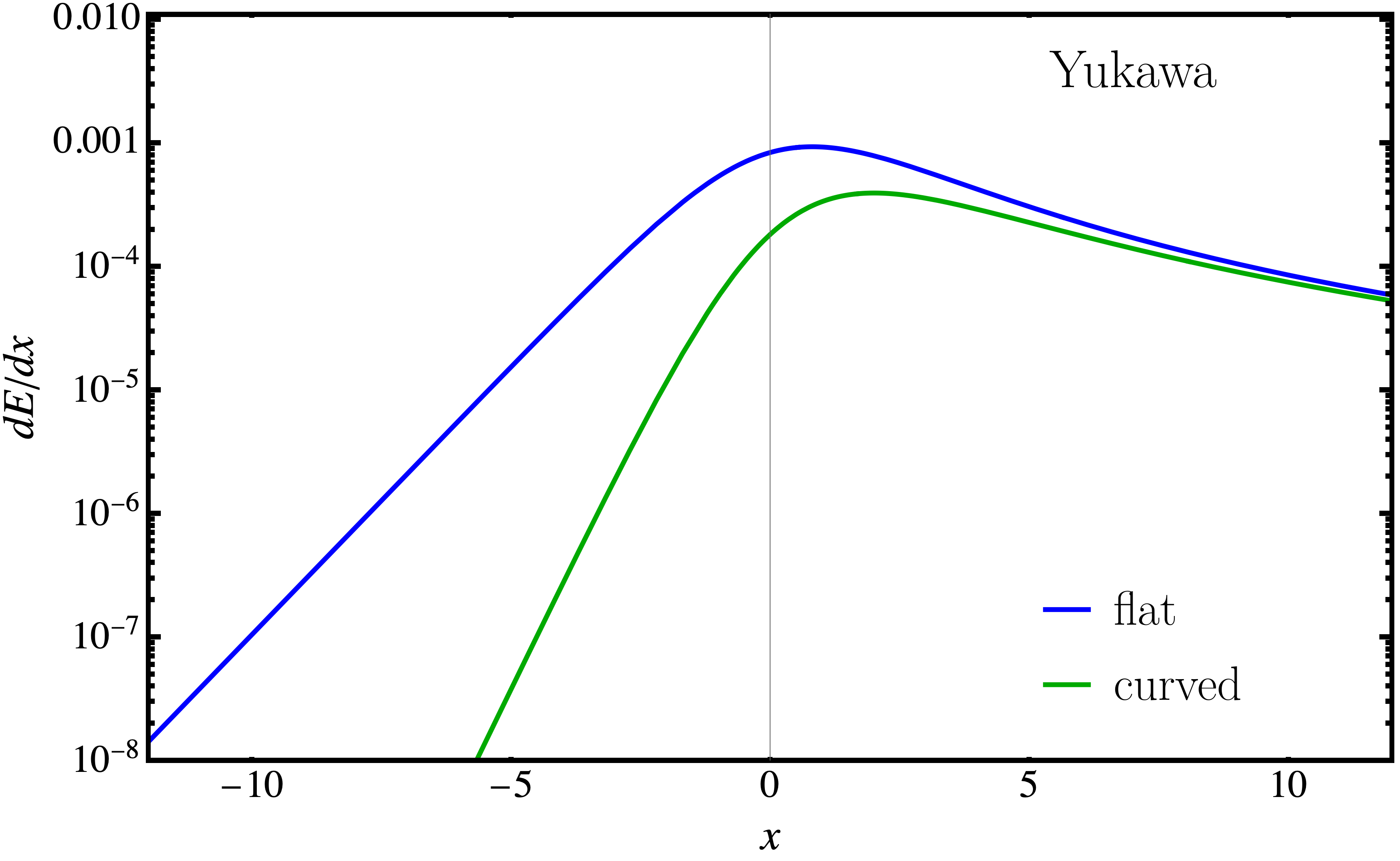}
    \caption{Yukawa.}
    \end{subfigure}
    \begin{subfigure}[]{0.495\textwidth}
        \includegraphics[width=\linewidth]{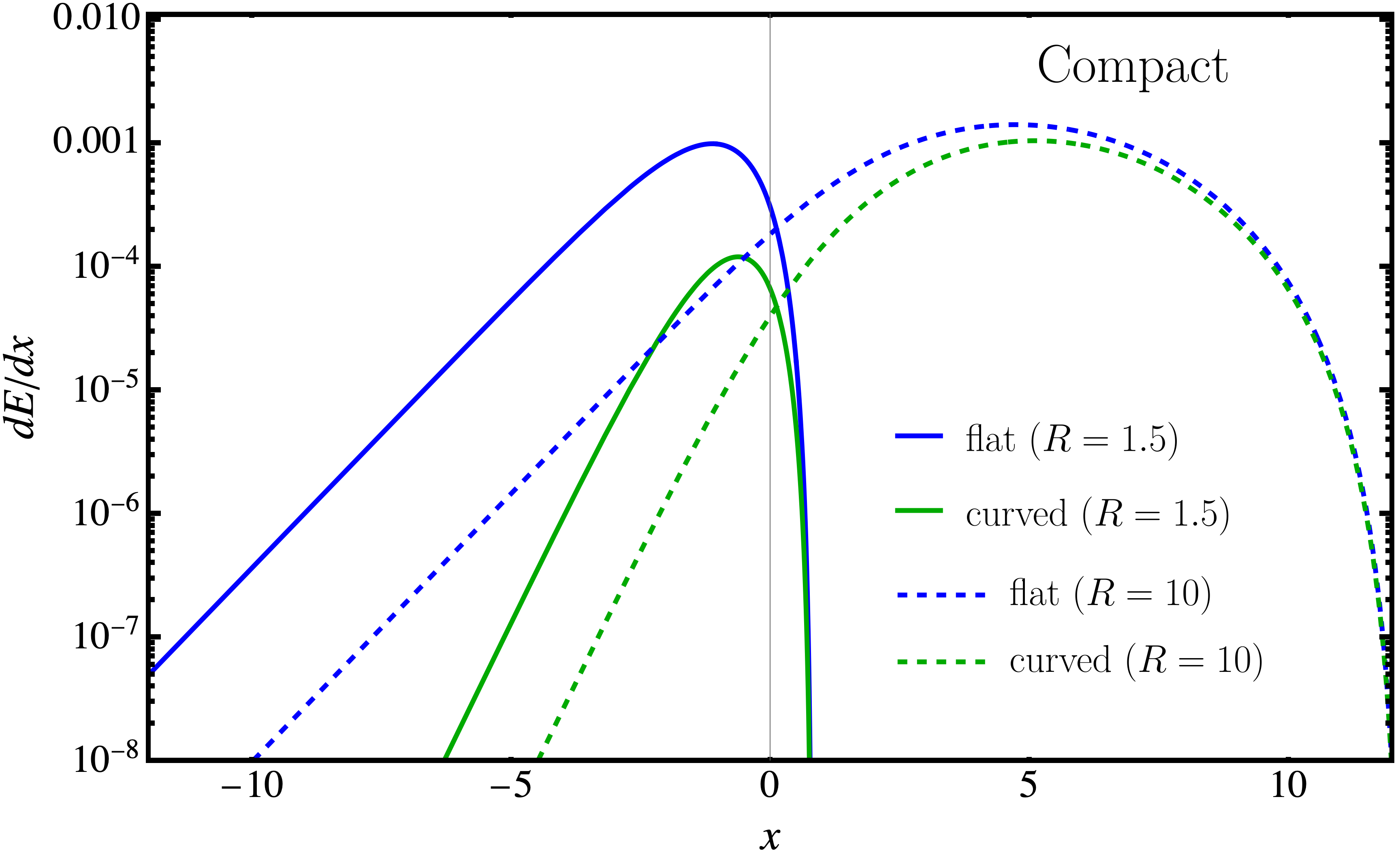}
    \caption{Compact.}
    \end{subfigure}
    \caption{Distribution of the energy as a function of the distance from the horizon in tortoise coordinates for an initial Yukawa and Compact profiles with $g_{Y,C}=1$. All quantities are expressed in units of $r_H$, which was fixed to $r_H=1$ for the plots.}
    \label{fig:differential-energy}
\end{figure}

The distribution of the energy as a function of the distance from the horizon in tortoise coordinates can be seen in Fig.~\ref{fig:differential-energy}.
The flat and curved profiles agree fairly well for $x\gtrsim 2$, and start to diverge significantly for $x \lesssim 0$, when $f(x)$ starts to approach the exponential behaviour. This is not surprising, given that this is the region where the gravitational potential is substantial.

\section{Results}
\label{sec:results}

Let us now turn to the numerical results for the various initial conditions and compare them to the flat space estimates. We will focus on the shape and qualitative features of the signal, its total energy, and its spectral composition.

Since any experimental observer will be located far away from the surroundings of the BH, we are interested in how much energy escapes from the BH once it is formed. To do so, we can take a finite (large) volume sphere of radius $x^*$ around the BH and calculate the energy flow going across its surface. We can then define the fraction of the energy that escapes as
\begin{equation}
\varepsilon=\frac{4\pi r(x^*)^2\int_0^\infty T^{0i}(x^*,t)dt}{4\pi\int_{-\infty}^{x^*}\rho(x,0)dx}\,,
\label{eq:energyration}
\end{equation}
where
$\rho(x,t)$ is the energy density\footnote{For convenience, in the definition of the energy density, we have effectively absorbed the usual $r(x)^2$ dependence arising in the integration over volume.} derived in Eq.~\eqref{eq:hamiltonian}
\begin{equation}
    \rho(x,t)=\frac{1}{2}\left(\dot{\psi}^2+\left(\psi'-\psi \frac{f(x)}{r(x)}\right)^2\right)\,.
    \label{eq:energydensity}
\end{equation}
In practice we compute this at a point far beyond the peak of the potential ($\hat{x}=0$) and after a long time where most of the field near the potential has relaxed to $0$ and the energy is concentrated in points where the potential is negligible (see, e.g., Fig.~\ref{fig:plotstatic1} for $\hat{t}=45$). In this way, to a good approximation, this energy will correspond to all the energy that escapes from the BH. Then we compare it with the total initial energy of the field. In~\cite{deGiorgi:2024pjb}, we naively estimated this ratio to be $\varepsilon\sim 0.5$, based on the assumption that the ingoing part of the wave gets totally absorbed, and the outgoing one can perfectly escape. We will comment on the validity of this reasoning case by case.

As also discussed in~\cite{deGiorgi:2024pjb}, the experimental sensitivity not only depends on the total energy released, but crucially also on the frequency spectrum. Given a signal function $h(t)$, we can define the Fourier Transform (FT) and the power spectral density~(PSD) as 
\begin{align}
  \label{eq:powerspe} &\Tilde{h}(\omega)=\int_{-\infty}^{\infty}h(t)e^{-i\omega t}dt~, &\text{PSD}[\Tilde{h}](f)=f |\Tilde{h}(f)|^2~,
\end{align}
where $f=\omega/(2\pi)$. 
In the following, we will consider
\begin{align}
    &h(t)=\dot{\phi}(t)\,, &\text{Qty. of interest}=\frac{(4\pi r^2) }{E}\times \text{PSD}[\dot{\phi}]\,,
    \label{eq:qoi}
\end{align}
where $E$ is the total energy that manages to escape.
This quantity has two advantages. First, the time derivative of the field often appears in experimental observables. Secondly, considering the derivative also avoids problems of convergence of the FT, which may arise in some of the static initial conditions. Note that, in the definition, it does not matter which $r$ is chosen as long as the quantity is computed sufficiently far away from the BH.

\subsection{Static initial conditions}
\subsubsection{Yukawa-like static}
Let us start with the simple static Yukawa-like distribution given in Eq.~\eqref{eq:initial-Y}. In our dimensionless field variables, this reads,
\begin{align} 
 &\psi_\text{IC}(\hat{x})=\frac{g_{\text{YL}}}{4\pi}\,,
 &\dot\psi_\text{IC}(\hat{x})= 0~.
\label{initialconditions}
\end{align}
Since the equation of motion is linear, we can set $g_{\text{YL}}=4\pi$ without loss of generality.  

\begin{figure}[t]
\centering
\centering
\begin{subfigure}[]{0.495\textwidth}
\includegraphics[width=\textwidth]{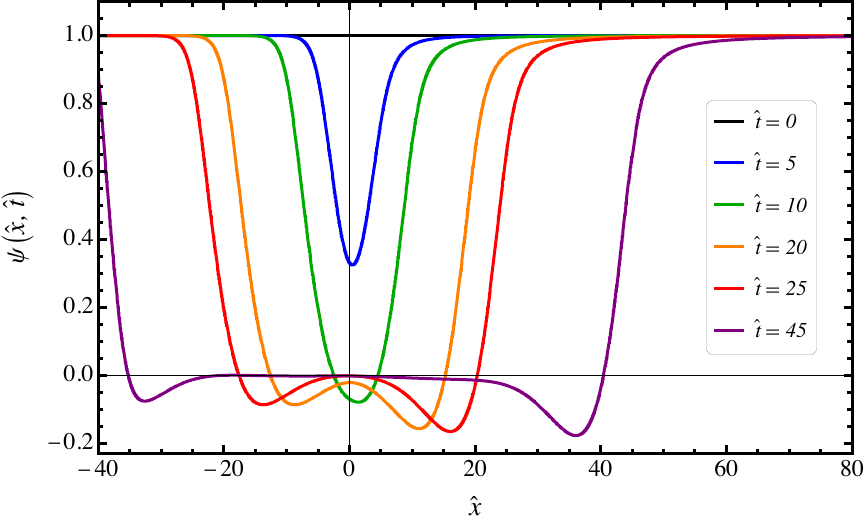} 
\caption{\label{fig:plotstatic1left}}
\end{subfigure}
\begin{subfigure}[]{0.495\textwidth}
\includegraphics[width=\textwidth]{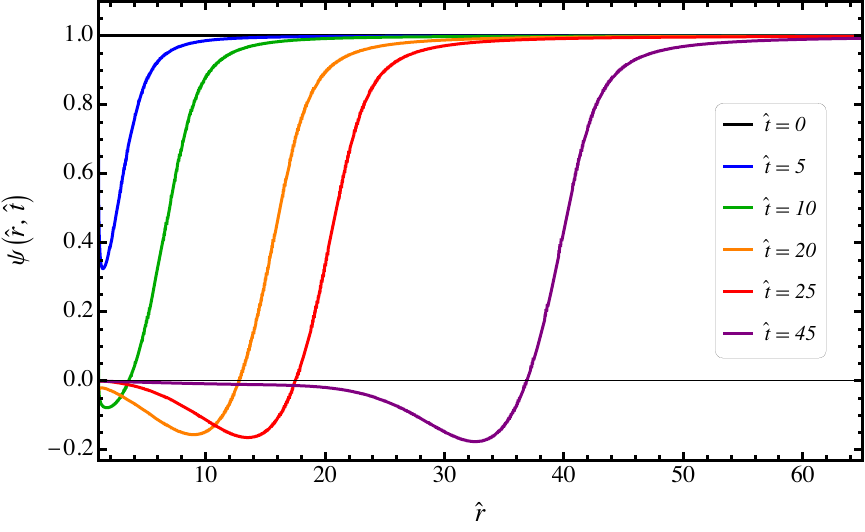}
\caption{\label{fig:plotstatic1right}}
\end{subfigure}
\caption{Spatial distribution of the field at different times for an initial Yukawa-like distribution. The left panel uses tortoise coordinates, whereas the right one uses Schwarzschild coordinates.}
\label{fig:plotstatic1}
\end{figure}

The field evolution in space in Schwarzschild and tortoise coordinates is shown in the two panels of Fig.~\ref{fig:plotstatic1}.
As we can see, for late times, the field eventually relaxes to zero everywhere. Although the potential extends over the whole space, its influence is mainly concentrated around $x\approx 0$. Accordingly, this is where the relaxation starts. 
For $\hat{x}\gg1$, i.e. at large distances from the black hole, tortoise coordinates match Schwarzschild coordinates, and in this region the shapes of the outgrowing wave front match. 

\begin{figure}[t]
\centering
\begin{subfigure}[]{0.485\textwidth}
\includegraphics[width=\textwidth]{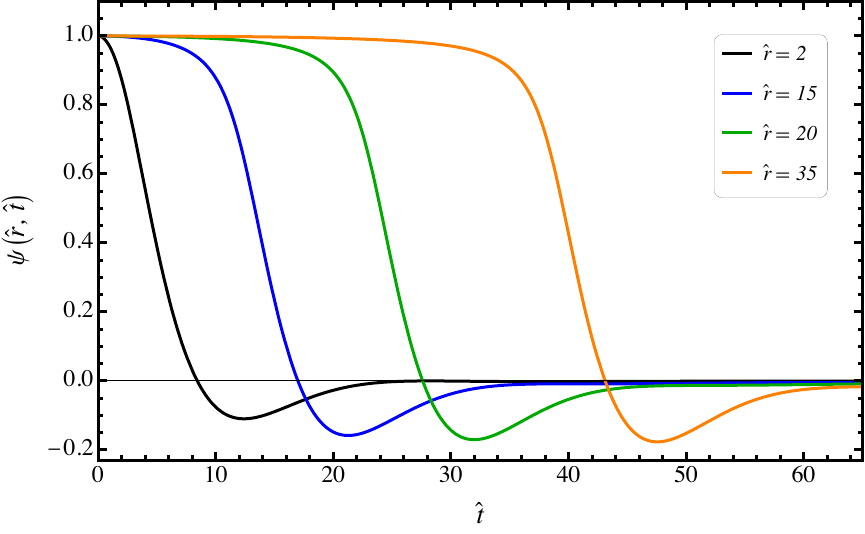} 
\caption{}
\label{fig:plotstatic3}
\end{subfigure}
\begin{subfigure}[]{0.505\textwidth}
\includegraphics[width=\textwidth]{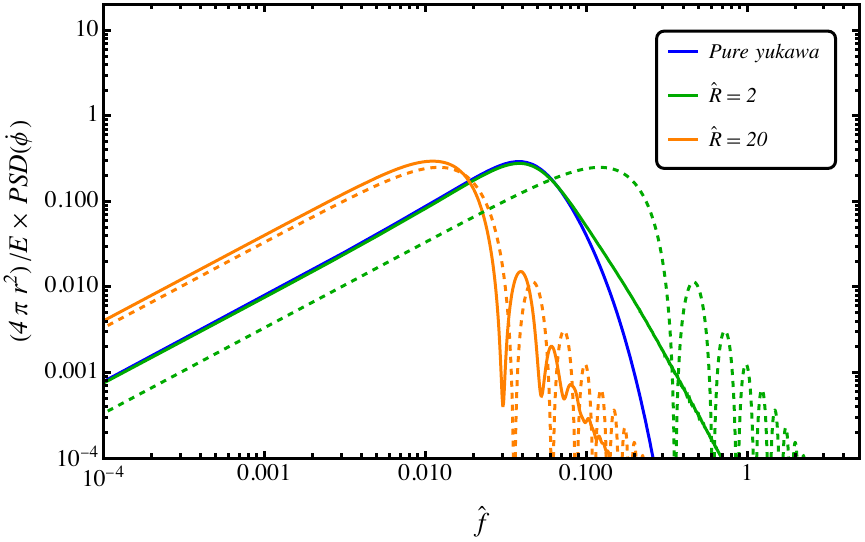} 
\caption{}
\label{fig:poweryuk}
\end{subfigure}
\caption{\textbf{Left:} Time evolution of the field at different space points for an initial Yukawa-like distribution. \textbf{Right:} Power spectral density for the Yukawa-like and different homogeneous charged spheres initial configurations. The curved and flat space calculations are shown with solid and dashed lines, respectively.}
\label{}
\end{figure}

In Fig.~\ref{fig:plotstatic3}, we show how the field evolves as a function of time at different fixed points in space.
In our modelling, the BH ``appears'' and the source ``disappears'' at  $\hat{t}=0$. In flat space, due to causality, the information should not arrive at a spatial point $\hat{r}$ until $\hat{t}= \hat{r}-1$. Note, however, that this is not strictly true in our modelling due to the curvature potential. The initial field configuration corresponds to the static solution with flat metric background, which is not a static solution of the Schwarzschild one. Therefore, the field becomes dynamical over a small region of space, even before the information of the source's disappearance has propagated. 
That said, far away from the BH this is likely to be a small effect as the Schwarzschild metric approaches the flat space one and the potential is very small.

As can be seen from the figure, the time evolution far from the BH looks pretty similar between the different points, only shifted in time. This is as expected, since at large distances the influence of the BH is negligible and the evolution should just be that of a massless field. Up to the $1/r$ decrease in amplitude, which is absorbed into the definition of $\psi$,  the massless field does not have any dispersion, and therefore the evolution preserves the shape.

Energy-wise, from our numerical solution, we find that
\begin{equation}
    \varepsilon\approx 0.52~,
\end{equation}
which means that approximately half of the energy of the field escapes the BH. This is (somewhat surprisingly) in agreement with the naive expectation of~\cite{deGiorgi:2024pjb}. We will elaborate more about deviations from this in Sec.~\ref{sec:compact-static}.

In Fig.~\ref{fig:poweryuk} we show, as the blue curve, the spectral quantity of interest defined in Eq.~\eqref{eq:qoi}. The spectrum exhibits a peak at the frequency of the first quasi-normal mode which in dimensionless units corresponds to $\hat{f}=fr_H\approx0.035$, that is the lowest mode that can be excited \cite{Berti:2009kk,Kokkotas:1999bd}.

\subsubsection{Homogeneous charged sphere}
Let us also consider an, initially static, homogeneously charged sphere. This gives the initial conditions
\begin{align}
    &\psi_{\text{IC}}(\hat{x})=\frac{g_{\text{YL}}}{4\pi}\times
\begin{cases} 
\dfrac{\hat{r}(\hat{x})}{2\hat{R}}\left(3-\dfrac{\hat{r}(\hat{x})^2}{\hat{R}^2}\right)\ & \hat{r}(\hat{x})<\hat{R}, \\[1mm]
1 & \hat{r}(\hat{x}) \ge \hat{R}~,
\end{cases}
     &\dot\psi_\text{IC}(\hat{x}) = 0~.
\end{align}

\begin{figure}[t]
\centering
\begin{subfigure}[]{0.495\textwidth}
\includegraphics[width=\textwidth]
{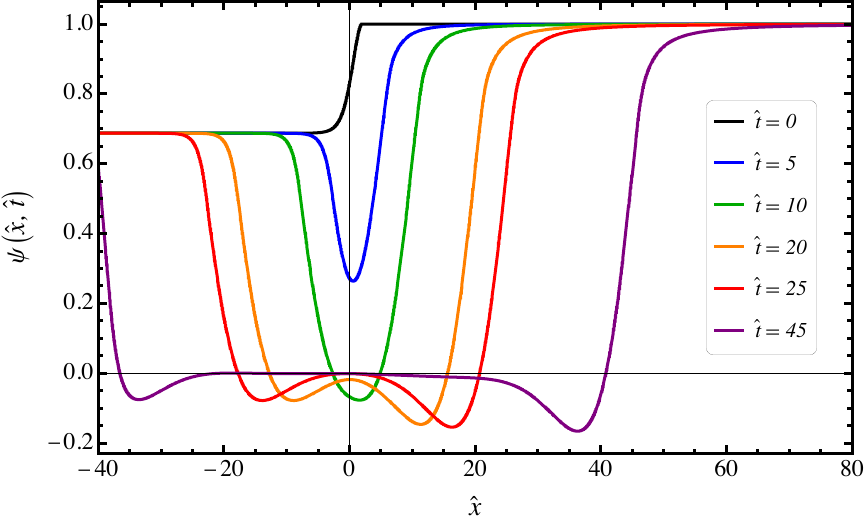} 
\caption{\label{fig:plothomogenenous1left}}
\end{subfigure}
\begin{subfigure}[]{0.495\textwidth}
\includegraphics[width=\textwidth]{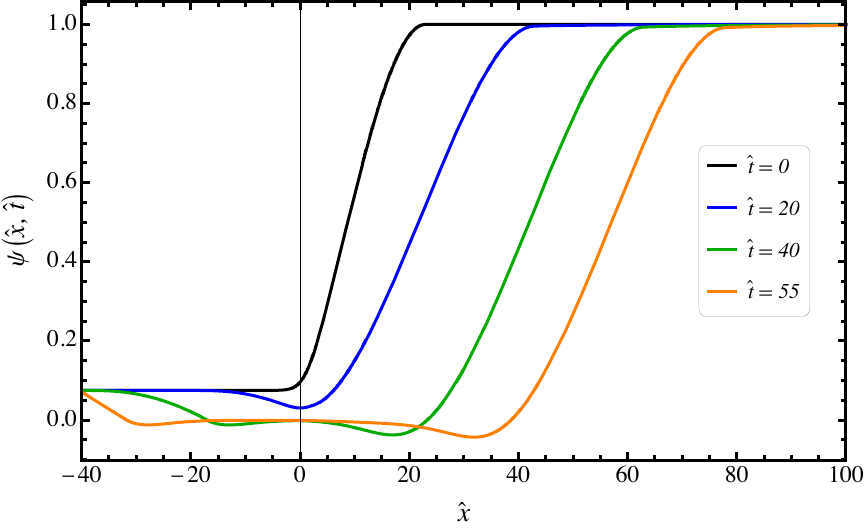}
\caption{\label{fig:plothomogenenous1right}}
\end{subfigure}
\caption{Spatial distribution of the field at different times for a homogeneous charged sphere of $\hat{R}=2$ (left panel) and $\hat{R}=20$ (right panel).}
\label{fig:plothomogenenous1}
\end{figure}

The field time evolution is shown in Fig.~\ref{fig:plothomogenenous1}.  
As we may expect, for relatively small values of $\hat{R}$, the time evolution, shown in Fig.~\ref{fig:plothomogenenous1left}, looks very similar to the Yukawa case. For a larger radius, the shape looks different, as can be seen in  Fig.~\ref{fig:plothomogenenous1right}. In this case, the size of the homogeneous sphere starts playing a role, and the shape of the propagating wave-front looks very similar to the naive flat space approximation presented in \cite{deGiorgi:2024pjb}.

From the numerical solution we can, again, compute the power spectral density using Eq.~\eqref{eq:qoi}. This is shown in Fig.~\ref{fig:poweryuk}. The behaviour of a small charged sphere (for example $\hat{R}=2$) looks very similar to the pure Yukawa initial configuration. They share the same peak at the first quasi-normal mode, and the behaviour for low frequencies is identical.  If we compare it with the flat space-time case (dashed line), the peak of the spectrum is redshifted, as we would have expected naively. Regarding the general shape, we have higher power for lower frequencies because low-frequency ingoing modes are almost perfectly reflected by the potential barrier (cf.~\cite{Das:1996wn,Kanti:2002nr} and also the discussion below). On the other hand, for higher frequencies, we have a suppression in the power compared to the flat case because they get almost fully transmitted through the barrier. 

For larger spheres (for example, $\hat{R}=20$), we will, in general, have a more similar spectrum to the flat one. Note, however, that the wiggles get damped in the curved case. We think that these wiggles in the flat space approximation come from the interference of high-frequency modes; in the GR case, they get suppressed due to the fact that the incoming part of high-frequency modes gets almost fully absorbed into the BH, leading to a suppression of the interference.

\subsubsection{Compact static}
\label{sec:compact-static}
As done in~\cite{deGiorgi:2024pjb} we can also assume a field configuration in which the energy is mostly confined to a finite region around the star, characterised by a large distance decay of the field faster than $1/r$. As a concrete example, we consider
\begin{align}
    &\psi_{\text{IC}}(\hat{x}) =
\begin{cases} 
\dfrac{g_C  \hat{r}(\hat{x})(\hat{R} - \hat{r}(\hat{x}))^3}{\hat{R}^4} & \hat{r}(\hat{x})<\hat{R}, \\[1mm]
0 & \hat{r}(\hat{x}) \ge \hat{R}~,
\end{cases}
     &\dot\psi_\text{IC}(\hat{x}) = 0~.
\end{align}
Here, $\hat{R}$ is the dimensionless radius within which all the energy is confined. Since the equation of motion is still linear, we choose $g_{\text{C}}=1$ without loss of generality.

\begin{figure}[t]
\centering
\begin{subfigure}[]{0.495\textwidth}
\includegraphics[width=\textwidth]{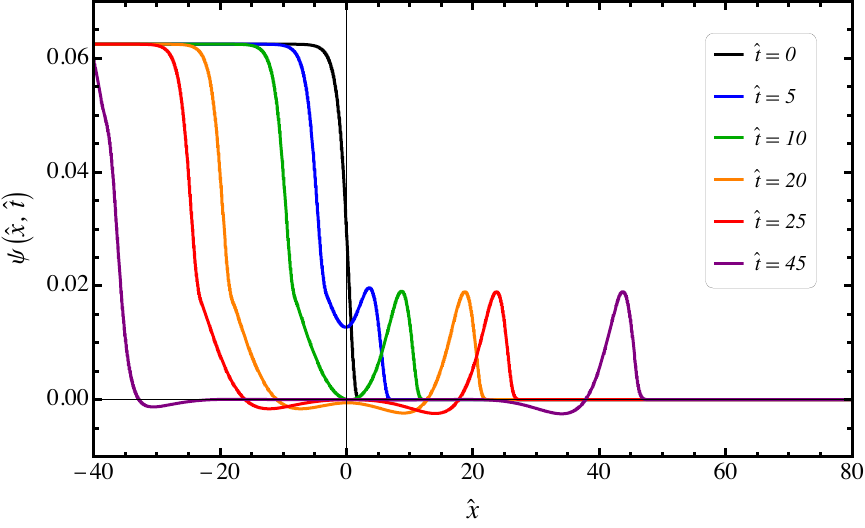} 
\caption{\label{fig:plotcompact2left}}
\end{subfigure}
\begin{subfigure}[]{0.495\textwidth}
\includegraphics[width=\textwidth]{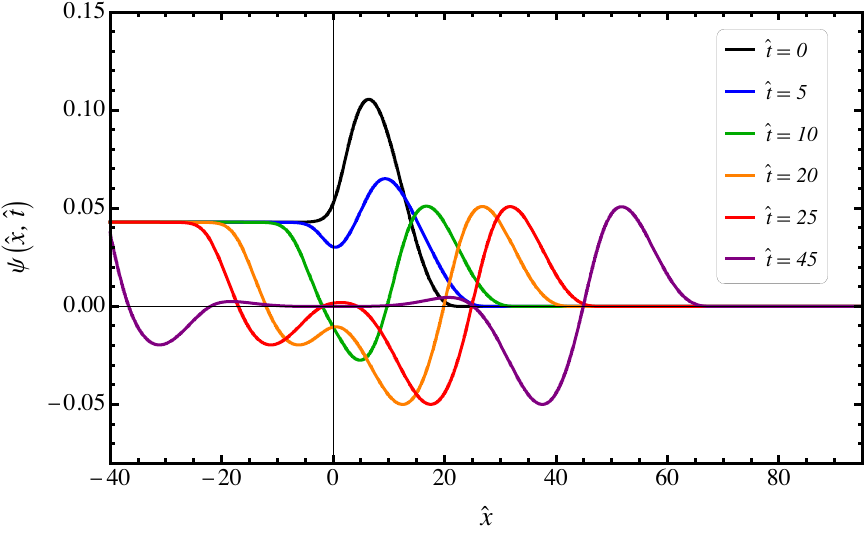}
\caption{\label{fig:plotcompact2right}}
\end{subfigure}
\caption{Spatial distribution of the field at different times for the compact static initial configuration for $\hat{R}=2$ (left) and $\hat{R}=20$ (right).}
\label{fig:plotcompact2}
\end{figure}

As an example, the evolution for $\hat{R}=2$ is shown in Fig.~\ref{fig:plotcompact2left}. As in the Yukawa-like and homogeneously charged spheres cases, the field relaxes to zero everywhere. 
However, its time evolution is quite different. In the case of a compact initial configuration, there are non-trivial contributions of ingoing and outgoing modes of the wave equation \eqref{eq:eomtortoisse}. At $t=0$ the field distribution divides into an ingoing and an outgoing front. Both wave fronts scatter in the potential, leading to the negative tails of the two wavefronts. Once the wave-fronts are far from the influence of the noticeable part of the potential, they propagate freely. The incoming front moves towards the Schwarzschild radius which, in tortoise coordinates, corresponds to $\hat{x}\rightarrow-\infty$ and will take infinite time to reach. On the other hand, the outgoing front escapes from the BH and propagates freely without the influence of the potential once it is far enough from the BH. As can be seen from the figure, this outgoing wave front has a tail that comes from the reflection of the potential as mentioned before. Some of the less energetic incoming modes are reflected by the potential, which changes the phase, leading to the negative tail that we can observe in the plot.

\begin{figure}
\centering
\begin{subfigure}[]{0.47\textwidth}
\includegraphics[width=\textwidth]{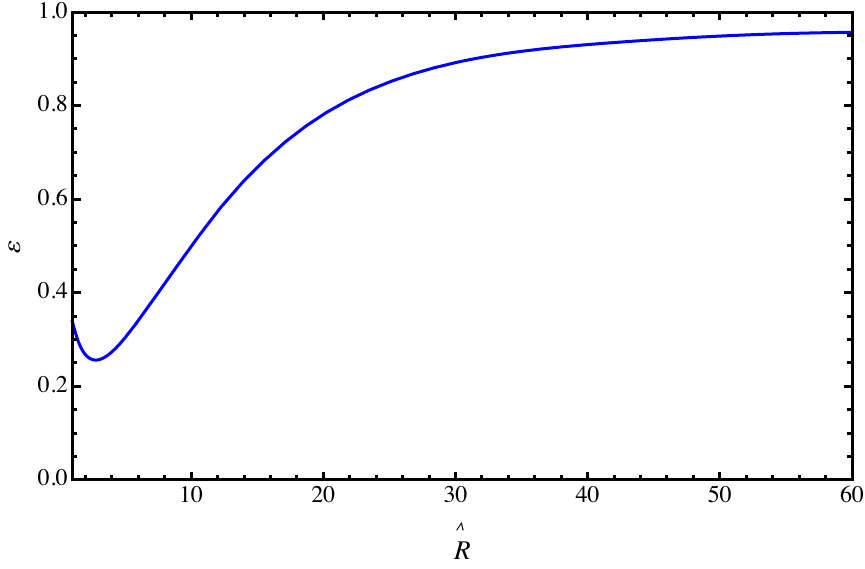} 
\caption{}
\label{fig:plotration}
\end{subfigure}
\begin{subfigure}[]{0.51\textwidth}
\includegraphics[width=\textwidth]{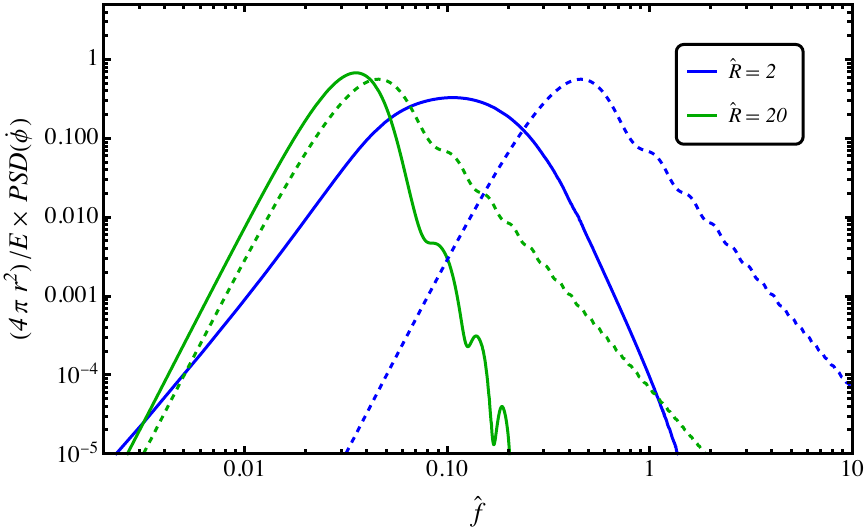} 
\caption{}
\label{fig:plotpower2}
\end{subfigure}
\label{}
\caption{\textbf{Left.} Fraction of the energy $\varepsilon$ that escapes (see Eq.~\eqref{eq:energyration}) as a function of the size $\hat{R}$ of the compact source. \textbf{Right.} Power spectral density for different-sized compact sources. GR calculation, solid and flat space approximation, dashed.}
\end{figure}

For wider initial configurations as shown in Fig.~\ref{fig:plotcompact2right} ($\hat{R}=20$), it can be seen that, although the general behaviour looks similar, the reflected tail is much bigger. This is what we expected, since for wider distribution we expect to excite lower frequency modes (see, e.g., Fig. \ref{fig:plotpower2}) that are more likely to be reflected at the potential barrier. Note that, the shape of the escaping front looks very similar to the flat case presented in \cite{deGiorgi:2024pjb}, which is what we expect for large sources.

As we did for the Yukawa configuration, we can compute the fraction of energy that escapes from the BH. This quantity, of course, depends on the size of the compact initial field configuration.
This is shown in Fig.~\ref{fig:plotration}.
Notably, this fraction is sizable, even for objects that are not much larger than the Schwarzschild radius. The reason for this is that, in this case, there is an outgoing and relatively high-frequency part of the field, which overcomes the potential barrier.
For large radii, the fraction actually approaches $1$. This is different to a, perhaps naive expectation of $1/2$, resulting from the energy being equally distributed between ingoing and outgoing waves. However, for large radii, the field is dominated by relatively low frequencies and the ingoing waves are nearly totally reflected. This is in line with the observation made in the calculation of grey body factors in black holes~\cite{Das:1996wn,Kanti:2002nr}.

\bigskip

The power spectral density is shown in Fig.~\ref{fig:plotpower2} for different-sized compact configurations (solid lines) alongside the naive flat space calculation (dashed lines).
We find that the spectrum is typically peaked at lower frequencies compared to the flat space case  (peaked around $\omega R\sim5$ \cite{deGiorgi:2024pjb}) due to the redshift caused by the gravitational field. This is crucial for the experiments as they typically are sensitive to certain frequency ranges.
For larger sources, the peak and the shape look more similar to the flat case. Note that, in general, we expect the power to be suppressed for higher frequencies compared to the flat case since high frequency in-going modes get almost fully totally transmitted trough the potential barrier. On the other hand, low frequencies get almost fully reflected on the potential barrier leading to enhanced power compared to the flat case.

\subsubsection{Spherical shell}
Another interesting scenario that can give us some insights into the physics is a spherical shell configuration given by,
\begin{align}
&\psi_{\text{IC}}(\hat{r}) =
\begin{cases}
0 & \hat{r} \le \hat{R}_1, \\[1mm]
\frac{ (\hat{r} - \hat{R}_1)^3 (\hat{R}_2 - \hat{r})^3}{(\hat{R}_2-\hat{R}_1)^6} & \hat{R}_1 < \hat{r} < \hat{R}_2, \\[1mm]
0 & \hat{r} \ge \hat{R}_2~,
\end{cases}
&\dot\psi_\text{IC}(\hat{x}) = 0~.
\end{align}

\begin{figure}[t]
\centering
\begin{subfigure}[]{0.495\textwidth}
\includegraphics[width=\textwidth]{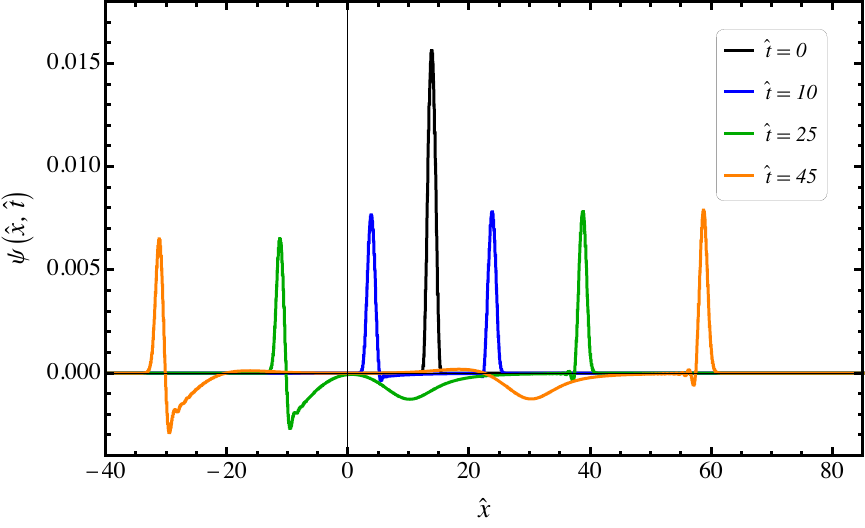} 
\caption{\label{fig:plotbump1left}}
\end{subfigure}
\begin{subfigure}[]{0.495\textwidth}
\includegraphics[width=\textwidth]{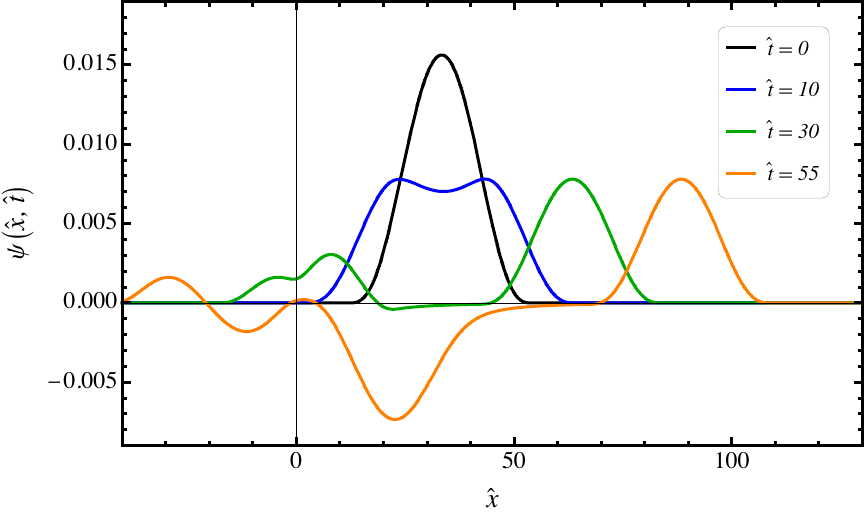}
\caption{\label{fig:plotbump1right}}
\end{subfigure}
\caption{Spatial distribution of the field at different times for a shell with $\hat{R_1}=10$ and $\hat{R_2}=13$ (left) and $\hat{R_1}=10$ and $\hat{R_2}=50$ (right).}
\label{fig:plotbump1}
\end{figure}

The initially vanishing time derivative of the field implies that, at the beginning, the initial shell will divide equally between incoming and outgoing modes. The resulting ``split'' in the shell can be seen in Fig.~\ref{fig:plotbump1} for small times. The outgoing part escapes the BH unperturbed since it is far from the influence of the potential. In contrast, the incoming part arrives at the potential wall, and some part of it gets transmitted while the other is reflected. 

\begin{figure}[t]
\centering
\includegraphics[width=0.48\textwidth]{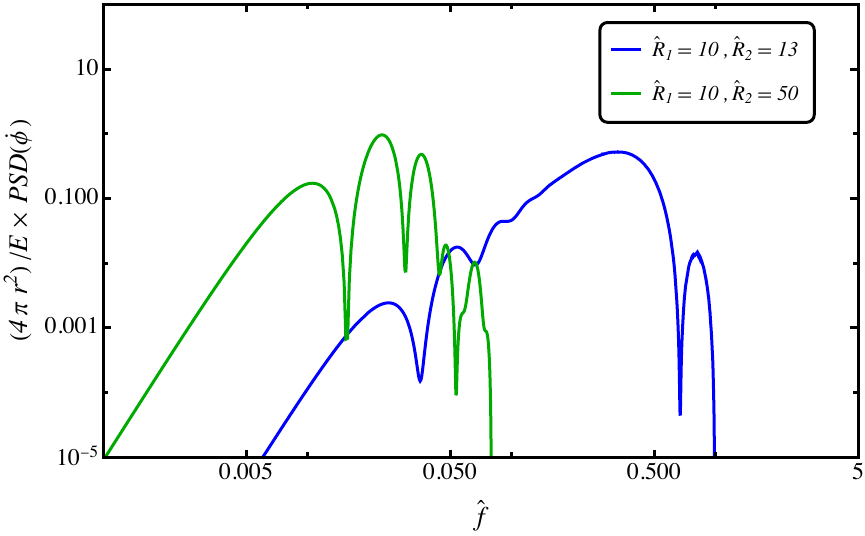} 
\caption{Power spectral density for the different shells. }
\label{fig:plotbumppower}
\end{figure}

Depending on the width of the shell, the potential reflects more or less of the incoming wave front. For $\hat{R_1}=10$ and $\hat{R_2}=13$ the fraction of energy that scapes is $\varepsilon\approx0.502$ and for $\hat{R_1}=10$ and $\hat{R_2}=50$ $\varepsilon\approx0.87$. For the outgoing modes, we expect all of them to escape the BH, since the initial bump is far from the influence of the potential wall. Hence, we expect $\varepsilon>0.5$. On the other hand, the incoming modes have a more interesting behaviour. Wider shells excite lower-frequency in-going modes, which are less energetic and, as noted before, are reflected more easily by the potential barrier, leading to a larger energy fraction escaping. This can also be seen in the power spectral density plotted in Fig.~\ref{fig:plotbumppower}. 

\subsection{Dynamical intial conditions}
\subsubsection{Yukawa-like ingoing}
So far, we have modelled the situation as if the black hole suddenly appeared out of nothing. In a more realistic situation, we would expect this to be preceded by some form of collapse of the initial configuration. It is intuitive to imagine that the field is dragged along in this collapse. As a first, simplistic approximation, let us consider an ``ingoing'' Yukawa-like distribution, i.e. where we only look at the part of the initial configuration that corresponds to ``ingoing'' waves. This can be expressed as an initial condition,
\begin{align}
 &\psi_\text{IC}(x)=1~,
 &\dot\psi_\text{IC}(x) =\psi'_\text{IC}(x)~.
\end{align}
Looking carefully at the initial conditions, we can already say that the field evolution will be identical to the Yukawa-like static. In fact, considering  $\psi_\text{IC}(x)=1$, automatically means that the spatial derivative vanishes, leading to  $\dot\psi_\text{IC}(x) =0$, which is exactly the static initial condition. 

We will consider a more physical approximation of the ``dragging along'' of the field configuration, below in Sec.~\ref{subsec:collapse}.

\subsubsection{Compact ingoing}
Another interesting situation consists of an ingoing compact distribution. Again, this can be relevant if the star rapidly collapses into a black hole, dragging the field along. Here, too, we consider a purely ingoing initial distribution given by
\begin{align}
    &\psi_{\text{IC}}(\hat{x}) =
\begin{cases} 
\dfrac{g_C  \hat{r}(\hat{x})(\hat{R} - \hat{r}(\hat{x}))^3}{\hat{R}^4} & \hat{r}(\hat{x})<\hat{R}, \\[1mm]
0 & \hat{r}(\hat{x}) \ge \hat{R}~,
\end{cases}
     &\dot\psi_\text{IC}(\hat{x}) = \psi'_\text{IC}(x)~.
\end{align}

\begin{figure}[t]
\centering
\includegraphics[width=0.48\textwidth]{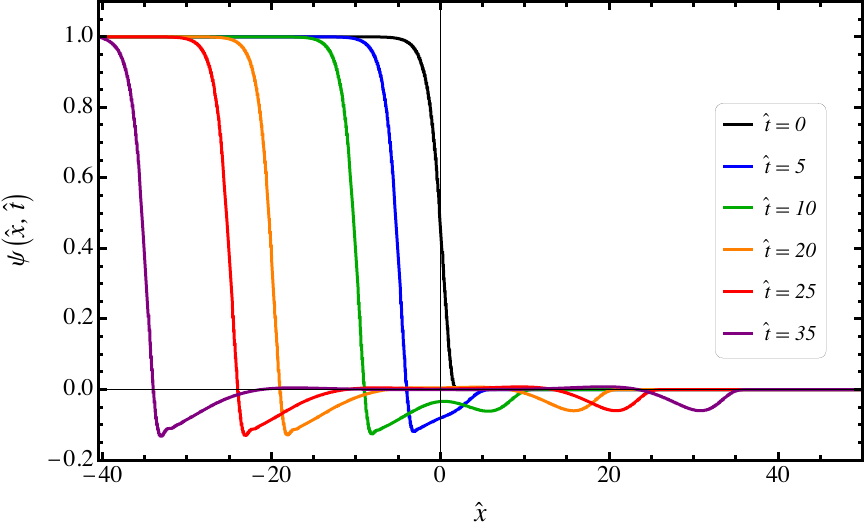} 
\caption{Spatial distribution of the field at different times for the compact ingoing initial configuration for $\hat{R}=2$.}
\label{fig:plotcompactswall}
\end{figure}

As can be seen in Fig.~\ref{fig:plotcompactswall}, even though the initial field distribution looks the same as the one in Fig. \ref{fig:plotcompact2left}, the time evolution of the field looks totally different. Having initially only ingoing modes, the only contribution that escapes the BH comes from the scattering of the incoming modes on the potential. We expect, however, that this energy is small since our initial field configuration is very close to $r_H$ and we are additionally only selecting the modes moving towards it. Using Eq.~\eqref{eq:energyration}, we find that, indeed, in this case $\varepsilon\approx0.005$, corroborating our intuition. Note, however, that this case is quite extreme and we will take a more realistic approach in the next section.

\subsubsection{Collapsing initial condition}\label{subsec:collapse}
Arguably, the initial conditions we have considered so far are not realistic with regard to what really happens when a BH forms. A star starts collapsing gravitationally, and once it is the size of the Schwarzschild radius, a black hole is formed. 

There are two aspects to model this process more realistically. Improving the metric and the field configuration. The metric outside the star is always Schwarzschild. Inside, it will take a more complicated form depending on the matter distribution inside the star and during the collapse. While this is an important effect, we will nevertheless focus on a better modelling of the initial condition for the field and continue to simply approximate the metric by a flat metric before the collapse and a Schwarzschild black hole one afterwards.

As an improvement for the initial condition, we use the field distribution from a homogeneous spherical Yukawa source contracting with a finite velocity in flat spacetime (see Eqs.~\eqref{eq:collapse1} and~\eqref{eq:collapse2} as well as App.~\ref{sec:contratingsource}). It is taken to disappear when the Schwarzchild radius is reached. Although this is still a significant simplification, as we are neglecting all general relativity effects when the star's radius is near the Schwarzschild radius, we believe it may capture some of the important features. 

\begin{figure}[t]
\centering
\begin{subfigure}[]{0.47\textwidth}
\includegraphics[width=\textwidth]{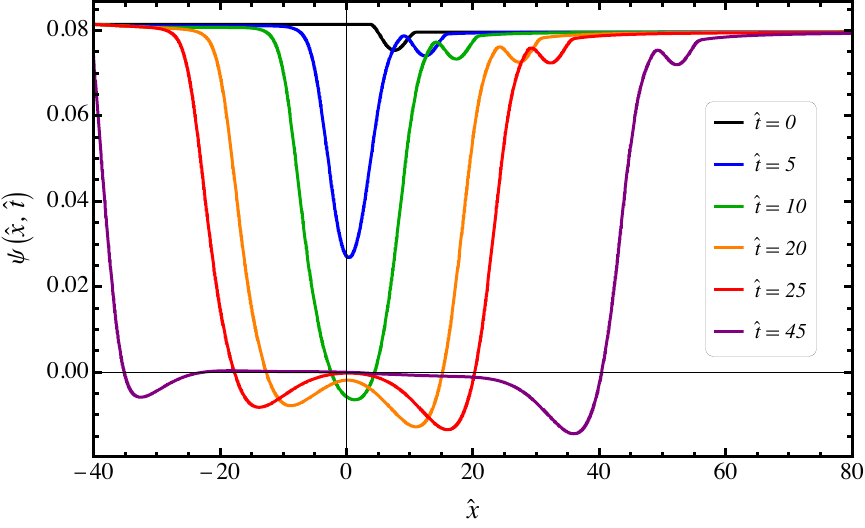} 
\caption{}
\label{fig:plotcollapsing1}
\end{subfigure}
\begin{subfigure}[]{0.52\textwidth}
\includegraphics[width=\textwidth]{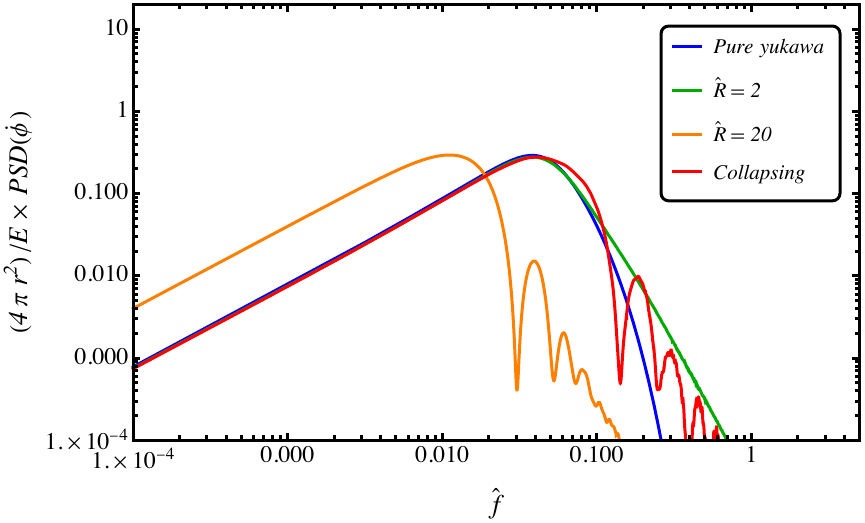} \caption{}
\label{fig:plotcollapsing2}
\end{subfigure}
\label{}
\caption{\textbf{Left.} Spatial distribution of the field at different times for the collapsing initial configuration, Eqs.~\eqref{eq:collapse1}, \eqref{eq:collapse2}, for $R_i=3r_H$ and $t_*=-6r_H$. \textbf{Right.} Comparison of the power spectral density for the collapsing initial condition and other previously considered initial field profiles.}
\end{figure}

In Fig. \ref{fig:plotcollapsing1}, we can see the time evolution for a source that has contracted from $\hat{r}=3$ to $\hat{r}=1$ in a time $\hat{t}=6$. As can be seen, the time evolution looks very similar to the static case we studied. This is because, even though the contraction is quite fast ($v=1/3$), the field distribution also reacts fast to the contraction, which is what we would expect for a massless field. The main observable difference compared to the static Yukawa initial condition is the small dip that can be observed in the field distribution. This feature comes from the contraction affecting how the field is distributed at $\hat{t}=0$.  Since the source is dragging the field along while it is contracting, we expect a bit more field near the Schwarzschild radius. This is indeed what we observe in the initial field distribution to the left of the dip compared to the right hand side of the dip. 

Since the initial conditions for the field are very similar to the static Yukawa-like case, we expect that the energy escaping the BH should be approximately the same. Applying Eq.~\eqref{eq:energyration}, we find that $\varepsilon\approx0.5$, leading to the conclusion that, again, approximately half of the energy escapes.

The power spectral density for this initial configuration can be seen in Fig.~\ref{fig:plotcollapsing2}. The shape looks similar to the pure Yukawa case. They share the peak at the first quasi-normal mode while differing at higher frequencies due to the dynamical initial conditions.

\section{Conclusions}
\label{sec:conclusions}

Even for rather feeble couplings of a new very light or massless scalar to individual matter particles, large matter objects, e.g. stars, may be surrounded by sizeable field configurations. Violent events such as supernovae or mergers may lead to black hole formation, the matter vanishes, and the field is effectively released from its binding coupling. It then propagates outwards in a Tsunami-like fashion, potentially inducing interesting transient signals in Earth-bound experiments. So far as discussed in~\cite{deGiorgi:2024pjb}.

However, it is justified to wonder if the strong gravitational field around the newly formed black hole has an effect on this release of the field configuration. What is the fraction of energy swallowed by the black hole? Are there noticeable changes in the experimentally important frequency spectrum of the emitted radiation? These are the questions motivating the present work. To address them, we (numerically) solve the field evolution equation in a Schwarzschild background and determine the field arriving at large distances from the black hole. We do so for various initial conditions motivated by the above scenario.

Turning first to the question of energy release. A very naive expectation is that roughly 50\% of the energy propagates outwards and the other 50\% effectively falls into the black hole. Indeed, we find that for static initial conditions, this is generally a good order of magnitude estimate even if gravitational effects are included. In this sense ``naive estimates are sometimes not as naive''. Interestingly, however, for many large initial objects (compared to the Schwarzschild radius), this expectation is even exceeded. This can be understood in tortoise coordinates, where the equation for the spherical waves essentially turns into a 1-d wave equation with a potential barrier somewhat outside the horizon. If the waves come from outside the barrier, there is a good chance of them being reflected, positively contributing to the energy release towards infinity. This is particularly pronounced for large field configurations dominated by lower frequency modes, which have a harder time penetrating the barrier. The same effect leads to a non-vanishing energy emission to infinity, even if we force the initial condition to essentially be ingoing into the black hole. If coming from outside the barrier, reflection takes place, leading to a smaller, but non-vanishing result.

Aside from the overall size of the field, experimental signals crucially depend on the frequencies contained in the transient wave. Here, a crucial effect is the expected redshift of the signal. Generally, the frequencies are smaller than in a naive flat space calculation. This effect is particularly pronounced for smaller (always compared to the Schwarzschild radius) objects. Depending on the type of experiment, this can have a significant impact on the sensitivity.

Finally, an important aspect is also the choice of a realistic initial condition. For this, we consider a shrinking homogeneously charged sphere. Using a realistic velocity for the shrinking process, we find only moderately small changes compared to an initially static charge configuration.

\section*{Acknowledgements}
We thank Ben Elder for providing access to his method of lines code, which was used in parts of the analysis presented in this work. We also thank Pierluca Carenza for helpful information and references on supernovae. YG is supported by the Australian Research Council’s Discovery Project (DP240103130) scheme.

\appendix
\section{Energy and conserved quantities}
\label{app:energy}

In this appendix, we briefly recall the main relations for the energy of the field in a gravitational background.

\subsection{Hamiltonian}
The Lagrangian density of a massless field incorporating gravitational effects is
\begin{equation}
    \mathcal{L}=\sqrt{-g}\, \frac{g^{\mu\nu}}{2}\partial_\mu\phi \partial_\nu \phi\,.
\end{equation}
The Hamiltonian density can be derived by performing the Legendre transformation, leading to
\begin{equation}
    \mathcal{H}=\sqrt{-g}\left[\, \frac{g^{00}}{2}\dot{\phi}^2-\frac{g^{11}}{2}(\phi')^2\right]\,,
\end{equation}
where we use $(x^0,x^1)=(t,r)$. Employing tortoise coordinates~(cf. Eq.~\eqref{eq:tortoise}), the Hamiltonian reads
\begin{equation}
    H=\int_V  \mathcal{H}=(2\pi)\int\limits_{-\infty}^\infty dx\, r(x)^2\,\left(\dot{\phi}^2+(\phi')^2\right)\,.
\end{equation}
Writing it in terms of $\phi(r)=\psi(r)/r$, it further simplifies to
\begin{equation}
    H=\int_V  \mathcal{H}=(2\pi)\int\limits_{-\infty}^\infty dx\, \,\left(\dot{\psi}^2+\left(\psi'-\psi \frac{f(x)}{r(x)}\right)^2\right)\,,
    \label{eq:hamiltonian-app}
\end{equation}
where $f(r)=1-\frac{r_H}{r}$ for a Schwarzschild metric as defined in Eq.~\eqref{eq:metric}.
The equation of motion in the same coordinates reads,
\begin{equation}
    \ddot{\psi}=\psi''-\psi\left(\frac{1}{r(x)}\frac{\partial f(x)}{\partial x}\right)=\psi''- V_s(x)\psi\,.
\end{equation}

Let us now try to understand how the energy flows through the system.
Integrating by parts and employing the EOM, one finds that the energy flow is determined only by the boundaries
\begin{equation}
    \frac{dH}{dt}=4\pi\left[\dot{\psi}\left(\psi'-\psi\frac{f(x)}{r(x)}\right)\right]_{x=-\infty}^{x=+\infty}\,.
\end{equation}
For any finite time, if $\dot{\psi}(x)$ vanishes at spatial infinity, as it should, one can safely set it to zero. The situation is less obvious at the horizon, i.e. $x=-\infty$. There we have,
\begin{equation}
\label{eq:energyflow}
    \frac{dH}{dt}=-4\pi\left.\dot{\psi}\left(\psi'-\psi\frac{f(x)}{r(x)}\right)\right|_{x=-\infty}=-4\pi\,r(x)^2\left.\dot{\phi}\,\phi'\right|_{x=-\infty}= -(4\pi r_H^2)\left.\dot{\phi}\,\phi'\right|_{x=-\infty}\,.
\end{equation}
The BH acts as a dissipative boundary term.

\subsection{Conserved quantity}
The \textbf{total} stress-energy momentum tensor is covariantly conserved during propagation
\begin{equation}
    T^{\mu\nu}_{;\mu}=0\,.
\end{equation}
For a Killing vector denoted by $X^\mu$, the following current is also conserved
\begin{align}
    &J^\mu = T^{\mu\nu}X_\nu\,, &J^\mu_{;\mu}=0\,.
\end{align}
The conserved charge reads
\begin{align}
    &Q=\int_V \sqrt{-g}J^0=\int_V \sqrt{-g}T^{0\nu}X_\nu\,, &\frac{dQ}{dt}=0\,.
\end{align}
In the Schwarzschild metric, a killing vector is $X^{\mu}=(1,\Vec{0})^\mu$\,. Therefore, the conserved charge reads
\begin{align}
    &Q=\int_V \sqrt{-g}T_{\phi+BH}^{00}g_{00}\,.
\end{align}
The energy-momentum tensor of a free radially propagating scalar field reads
\begin{align}
    &T^{\mu\nu}=\phi^{,\mu}\phi^{,\nu}-\frac{g^{\mu\nu}}{2}\phi^{,\alpha}\phi_{,\alpha}\,, &T^{00}=\frac{g^{00}}{2}\left(g^{00}(\dot{\phi})^2-g^{xx}(\phi')^2\right)
\end{align}
and therefore we can define in analogy
\begin{align}
    &Q_\phi=\frac{1}{2}\int_V \sqrt{-g}\left(g^{00}(\dot{\phi})^2-g^{xx}(\phi')^2\right)=(2\pi)\int_V dx\, r(x)^2 \left((\dot{\phi})^2+(\phi')^2\right)=H\,.
\end{align}
Notice that $Q_\phi$ by itself is not conserved; the conserved quantity is rather
\begin{align}
    &Q=Q_\phi +Q_\text{BH}\,, &\frac{dQ_{BH}}{dt}=-\frac{dQ_\phi}{dt}=(4\pi r_H^2)\left.\dot{\phi}\,\phi'\right|_{x=-\infty}\,.
\end{align}
The energy lost is the one that falls into the BH.

\section{Comments on the growth of the black hole and the shedding of scalar hair}\label{app:nohair}
In the case of a static Yukawa-like initial condition such as the one shown in Fig.~\ref{fig:plotstatic1left} (as well as for more general static initial conditions), we notice that the field $\psi$ at the horizon ($x\to-\infty$) does not change, i.e.
$\dot{\psi}(r_{H})=0$. 
Therefore, a non-vanishing field seemingly remains non-vanishing for all times. For a Yukawa-like initial condition, this is easy to see from the equation of motion, Eq.~\eqref{eq:motion}, in tortoise coordinates. For the Yukawa-like initial condition, the field value is spatially constant, hence $\psi''=0$. Moreover, the potential also vanishes near the horizon. Therefore, the time derivative of the field remains $0$. Indeed, any effect from switching on a potential that is essentially localized around $x\sim 0$ and that would cause the field to move away from this constant value needs an infinite time to reach the horizon located at $x\to-\infty$.

One may wonder how this observation fits with the picture that a black hole should not have any scalar features~\cite{Israel:1967wq,Israel:1967za,Carter:1971zc,Ruffini:1971bza,Carter:2009nex,Robinson:1975bv,Carter:1979wef,Mazur:2000pn} and whether there is something seriously wrong with the approximation. Let us therefore briefly discuss how the shedding of hair would happen in this situation in a slightly improved approximation.

\begin{figure}[t]
\centering
\includegraphics[width=0.48\textwidth]{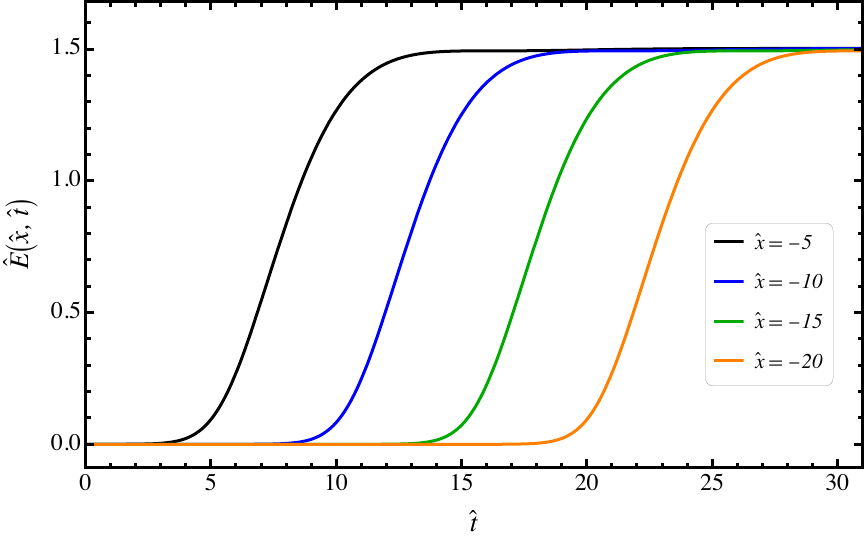} 
\caption{Evolution of the energy enclosed ($\hat{E}=E_{\text{enc}}(\frac{4\pi}{g_\text{Y}})^2r_H$) in spheres of different radii around the BH for the Yukawa-like initial field configuration.}
\label{fig:enery1}
\end{figure}

From Fig.~\ref{fig:plotstatic1} we notice that while the field remains constant at the horizon, it evolves to zero in the vicinity. Indeed, for longer times, the field is approximately zero, ever closer to the horizon.
In the vicinity of the horizon, we therefore have a large and increasing field gradient. This contains a significant amount of energy.
To quantify how the energy gets accumulated near the horizon, we can compute the time evolution of the energy enclosed in spheres of different radii around the BH, which we can calculate as
\begin{equation}
  E_{\text{enc}}(x,t)=4\pi\int_{-\infty}^{x}\rho(x',t)dx'~,
\label{eq:enclosed}
\end{equation}
where $x$ is the radius of the corresponding sphere in tortoise coordinates. As can be seen in Fig.~\ref{fig:enery1}, if we take, for example, the Yukawa initial field condition, the energy for the different spheres increases when the wave-front arrives and remains constant after that. This means that the energy is travelling towards the Schwarzschild radius which, in tortoise coordinates, will take infinite time to reach.

Having an accumulation of energy in the vicinity of the horizon eventually requires the horizon to grow, hence the seizable field values are ``swallowed''. Along these lines, the field values eventually approach zero near the (somewhat expanded) horizon.
Consequently, the long-term solution does not feature any scalar field value at the horizon. It has shed, or more precisely, ``eaten'' its hair.

The growth of the horizon is not included in our present approximation. However, it is an effect that happens at somewhat late times and does not significantly interfere with the evolution of the part of the field that is emitted from our initial source. Indeed, in the figures (e.g. Fig.~\ref{fig:plotstatic1}) we can see that the two parts of the field, one moving towards and the other away from the horizon, are well separated.
Moreover, as long as the total energy contained in the field is small compared to the mass of the formed black hole, the growth of the horizon is also expected to be modest, hence having little effect on the outgoing part of the field.

\section{Contracting source}
\label{sec:contratingsource}
For a somewhat more realistic modelling of the initial condition, let us consider a spherical source of radius $R(t)$ shrinking in time in flat space-time. For a homogeneous sphere of radius $R(t)$, the source current can be written as
\begin{align}
   &(\Box+m_\phi^2)\phi=J(r,t)\,, &J(r,t)=\dfrac{3g_Y}{4\pi R(t)^3}\Theta(-t)\Theta(R(t)-r)\,.
\end{align}
The radius $R(t)$ is a function of time and shrinks for $t<0$ until, eventually, the source disappears at $t=0$. The solution of the field in the $m_\phi=0$ limit, obtained via integration with the Green's function, is given by
\begin{equation}
    \phi(r,t)=\dfrac{3g_Y}{2\pi^2r}\int\limits_0^\infty d{k}\,\sin(kr)\int\limits_{-\infty}^{\min(0,t)}d\tau\,\sin(k(t-\tau))F[k R(\tau)]\,e^{\epsilon \tau}\,.
\end{equation}
Here, $\epsilon>0$ regularises the integral for $\tau\to-\infty$ and should be taken to zero after the integration. Moreover, we define
\begin{equation}
    F(kR)\equiv\frac{1}{(kR)^2}\left[\dfrac{\sin(k R)}{kR}-\cos{(kR)}\right]\approx \frac{1}{3}+\mathcal{O}(kR)\,.
\end{equation}
We now assume that, at time $t_\star<0$, the collapse happens and goes on until $t=0$, when the source disappears. We have three regimes:
\begin{enumerate}
    \item The source is static for $t<t_\star<0$ and $R=R_i$ is constant. The integration over time can be performed, and leads to
    \begin{equation}
        \phi(r,t<t_\star)=\dfrac{3g_Y}{2\pi^2}\int\limits_0^\infty d{k}\,\frac{\sin(kr)}{kr}\,F[k R_i]\,.
    \end{equation}
    As we are interested in the field outside the source, for $r>R_i$ the integration leads to the usual $1/r$ profile
    \begin{equation}
        \phi(r>R_i,t<t_\star)=\frac{g_Y}{4\pi r}\,.
    \end{equation}
    \item The source shrinks $t_\star\leq t<0$ from $R_i$ to $R_f$.  The result depends on the exact shape of $R(t)$. By first performing the integral over $k$ one gets for $r>R$
    \begin{equation}
       \phi(r,t)=\frac{3g_Y}{32\pi r} \int\limits_{-\infty}^{\min(0,t)}d\tau\,\frac{(r + R - t + \tau)|r - R - t + \tau| - 
       (r - R - t + \tau)|r + R - t + \tau|)}{R^3}\,,
    \end{equation}
    where above $R=R(\tau)$.
    \item The source disappears at $t=0$.
\end{enumerate}

For $t_\star<t<0$, the source shrinks.
In this time interval, we consider in the main text a linearly shrinking source from the initial radius $R_i$ to the final one $R_f$
\begin{equation}
\label{eq:decrease-linear}
    R(t)=R_i+\Theta(t-t_\star)(R_f-R_i)\left(1-\frac{t}{t_\star}\right)\,.
\end{equation}
With this ansatz, one can check that emission of the field takes place only when the source size accelerates, which corresponds to kicks at $t=t_\star$ and $t=0$. For $t_\star<t<0$, one has instead $\ddot{R}=0$. Consequently, it can happen that the time derivative of the field at the final size vanishes, $\dot{\phi}(r=R_f,t=0)=0$.
This does not always happen; it requires the shrinking process to be much slower compared to the relative initial and final sizes of the source, so that the field has enough time to be emitted and become static again. Let us define the shrinking velocity $\beta$ and the ratio of sizes, $\alpha$
\begin{align}
   &\alpha\equiv\frac{R_f}{R_i}\leq 1\,, &\beta\equiv  -\frac{\Delta R}{\Delta t}\leq 1\,.
\end{align}
Then, \textit{no-effect} at $r=R_f$ at $t=0$ is observed if
\begin{equation}
    \Delta t-R_i\geq R_f\,,
\end{equation}
which is equivalent to velocities smaller than
\begin{equation}
    \beta\leq \frac{1-\alpha}{1+\alpha}\,.
\end{equation}
Notice that this description holds for a contraction of the form as in Eq.~\eqref{eq:decrease-linear}. If the time dependence of $R(t)$ features non-vanishing acceleration, one expects also a non-vanishing time derivative of $\phi$.


\section{Validation of numerical results}\label{app:extrafig}

\begin{figure}[t]
\centering
\begin{subfigure}[]{0.48\textwidth}
\includegraphics[width=\textwidth]{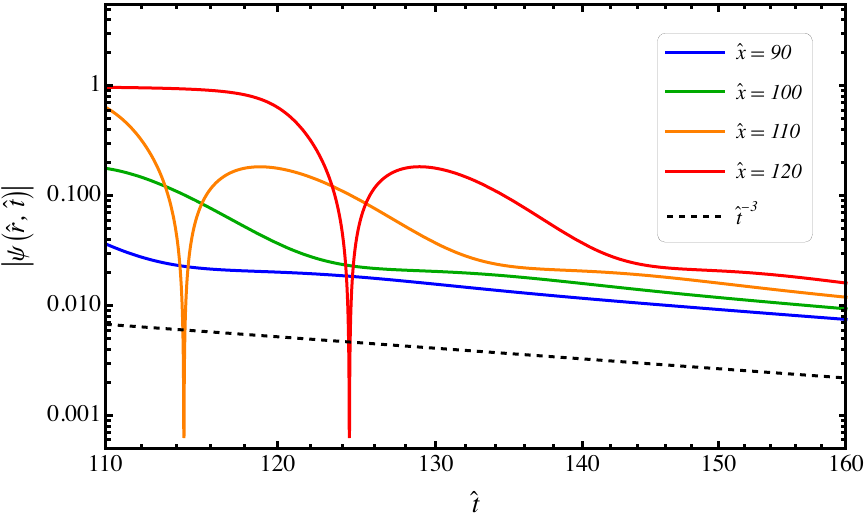} 
\caption{\label{fig:plotasymp1}}
\end{subfigure}
\begin{subfigure}[]{0.48\textwidth}
\includegraphics[width=\textwidth]{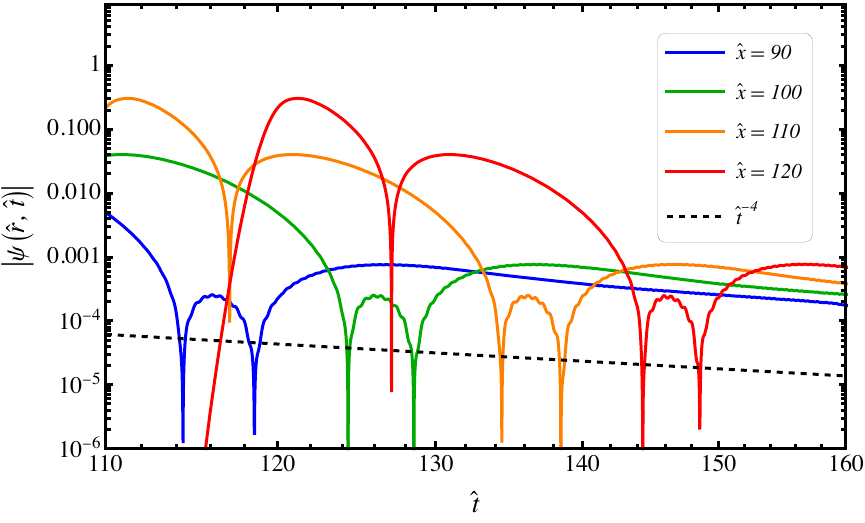} 
\caption{\label{fig:plotasymp12}}
\end{subfigure}
\caption{\textbf{Left.} Asymptotic behaviour of the field for a static Yukawa field configuration \textbf{Right.} Asymptotic behaviour of the field for a static compact field configuration with $\hat{R}=2$.}
\label{fig:plotasympcomb}
\end{figure}

A consistency check of our numerical framework is provided by the late-time behaviour of the field. At sufficiently late times, the field is expected to exhibit a power-law decay arising from backscattering off the long-range tail of the spacetime curvature potential, as mentioned in Sec.~\ref{sec:grevolution}. This phenomenon is commonly referred to as Price’s law and has been extensively studied in the literature \cite{Price:1971fb,Ching:1995tj,Donninger:2009rc,Donninger:2009zf}. The precise decay exponent depends on the nature of the initial perturbation. In particular, for initially static configurations, the scalar field decays asymptotically as $t^{-3}$ for a Yukawa-like profile, while the compact initial configuration leads to a steeper $t^{-4}$ fall-off.

To validate the reliability of our numerical solutions, we have explicitly verified that the late-time scaling of the field is consistent with these analytical expectations. This can be seen in Figs.~\ref{fig:plotasymp1} and~\ref{fig:plotasymp12}.


\bibliographystyle{JHEP}
\bibliography{references}
\end{document}